\documentclass[a4paper,11pt]{article}
\pdfoutput=1
\usepackage{amssymb,amsmath,amsfonts,makeidx,placeins,pbox,multirow}
\usepackage{graphicx,rotate,subcaption,color,slashed,cite,caption,epstopdf,verbatim}
\usepackage{longtable,tabu}
\usepackage{array}
\usepackage{color}
\usepackage{url}
\usepackage[table]{xcolor}
\usepackage{amsmath}
\usepackage{mathtools}
\usepackage{amssymb}
\usepackage[colorlinks=true,
            linkcolor=blue,
            citecolor=blue,
            urlcolor=blue]{hyperref}
\usepackage[normalem]{ulem}
\usepackage{graphicx}
\usepackage{svg}
\usepackage[justification=centering]{caption}
\usepackage{subcaption}
\usepackage{booktabs}
\usepackage{makecell}  
\newcolumntype{P}[1]{>{\centering\arraybackslash}p{#1}}
\usepackage{tablefootnote}

\def\lapp{\mathrel{\rlap{\raise.5ex\hbox{$<$}}
                    {\lower.5ex\hbox{$\sim$}}}}
\def\gapp{\mathrel{\rlap{\raise.5ex\hbox{$>$}}
                  {\lower.5ex\hbox{$\sim$}}}}

\usepackage{color}
\long\def\/*#1*/{}
\usepackage{color}
 \usepackage[normalem]{ulem}
\definecolor{darkgreen}{cmyk}{1,0,1,0.4}
\definecolor{darkred}{cmyk}{0,1,1,0.4}
\definecolor{rosso}{cmyk}{0,1,1,0.4}
\definecolor{rossos}{cmyk}{0,1,1,0.55}
\definecolor{rossoc}{cmyk}{0,1,1,0.2}
\definecolor{blu}{cmyk}{1,1,0,0.3}
\definecolor{blus}{cmyk}{1,1,0,0.6}
\definecolor{bluc}{cmyk}{1,1,0,0.1}
\definecolor{verde}{cmyk}{0.92,0,0.59,0.25}
\definecolor{verdec}{cmyk}{0.92,0,0.59,0.15}
\definecolor{verdes}{cmyk}{0.92,0,0.59,0.4}
\definecolor{grigio}{cmyk}{0,0,0,0.07}
\definecolor{rosa}{cmyk}{0,0.1,0.1,0.02}
\definecolor{rosino}{cmyk}{0,0.05,0.05,0.02}
\definecolor{rosas}{cmyk}{0,0.3,0.25,0.05}
\definecolor{celeste}{cmyk}{0.1,0,0,0.02}
\definecolor{giallino}{cmyk}{0,0,0.4,0.02}
\definecolor{rosso}{cmyk}{0,1,1,0.4}
\definecolor{rossos}{cmyk}{0,1,1,0.55}
\definecolor{rossoc}{cmyk}{0,1,1,0.2}
\definecolor{blu}{cmyk}{1,1,0,0.3}
\definecolor{bluc}{cmyk}{1,1,0,0.1}
\definecolor{blucc}{cmyk}{0.7,0.5,0,0}
\definecolor{viola}{cmyk}{0,1,0,0.6}
\definecolor{viola2}{cmyk}{0,1,0.2,0.6}
\definecolor{verde}{cmyk}{0.92,0,0.59,0.25}
\definecolor{verdec}{cmyk}{0.92,0,0.59,0.15}
\definecolor{verdes}{cmyk}{0.92,0,0.59,0.4}
\definecolor{verdino}{cmyk}{0.12,0,0.09,0.05}
\definecolor{giallo}{cmyk}{0,0,1,0}
\definecolor{gialloverde}{cmyk}{0.44,0,0.74,0}

\evensidemargin=\oddsidemargin
\def\bar {\overline}

\def\bea {\begin{eqnarray}}
\def\eea {\end{eqnarray}}

\def\beq{\begin{equation}}
\def\eeq{\end{equation}}
\def\barr{\begin{array}}
\def\earr{\end{array}}

\def\beq{\begin{equation}}
\def\eeq{\end{equation}}

\newcommand{\nc}{\newcommand}

\nc{\hi}{H}
\nc{\hit}{\widetilde{H}}
\nc{\hij }{\mbox{${\hi^\dag i\,\raisebox{2mm}{\boldmath ${}^\leftrightarrow$}\hspace{-4mm} D_\mu\,\hi}$}}
\nc{\hijt}{\mbox{${\hi^\dag i\,\raisebox{2mm}{\boldmath ${}^\leftrightarrow$}\hspace{-4mm} D_\mu^{\,a}\,\hi}$}}

\def\gev{\rm GeV}
\def\tev{\rm TeV}

\def\gev{\,\ensuremath{\mathrm{Ge\kern -0.1em V}}}
\def\tev{\,\ensuremath{\mathrm{Te\kern -0.1em V}}}

\DeclareUnicodeCharacter{2212}{-}
\UseRawInputEncoding

\begin{document}

\begin{center}
{\Large {\bf RG evolution and effect of intermediate new physics on $\Delta B=2$ six-quark operators}} \\
\vspace*{0.8cm} {\sf Mathew Thomas Arun \footnote{mathewthomas@iisertvm.ac.in}, Shyam M\footnote{shyam.m.work.27@gmail.com}, Ritik Pal\footnote{ritik24@iisertvm.ac.in }} \\
\vspace{10pt} {\small } {\em  School of Physics, Indian Institute of Science Education and Research, Thiruvananthapuram 695551, Kerala, India}
\normalsize
\end{center}
\bigskip
\begin{abstract}
The recent observation of possible 11 neutron-antineutron ($n$-$\bar{n}$) oscillation candidates with an expected background of $9.3\pm2.7$ events at Super-Kamiokande has renewed the interest in $\Delta B = 2$ transitions. In this work, we analyze the Renormalization Group (RG) running of mass dimension-9 six-quark operators, in $\bar{MS}$ scheme, that generate processes like $nn\to \pi^0\pi^0$, deuteron decay, $n$-$\bar{n}$ oscillations etc, evolving them from the electroweak scale to baryon number violation (BNV) scale ($\mathcal{O}(10^3~\text{TeV})$). Our goal is to systematically account for the influence of potential new physics at intermediate energies ($\gtrsim \mathcal{O}(10~ \text{TeV})$), especially given the fact that {\it Large Hadron Collider} has not ruled out new physics beyond $\sim 10~\text{TeV}$. To comprehensively investigate their influence, we consider two scenarios: (i) a minimal setup with only Standard Model degrees of freedom up to the high scale at $\mathcal{O}(10^3~\text{TeV})$, and (ii) an extended framework involving scalar and vector bosons above $\sim 10~\text{TeV}$ up till BNV scale. To facilitate further studies, we also provide a Python script\footnote{In https://github.com/rp-winter/bnv-running} that performs RG evolution of the BNV Wilson coefficients in the presence of generic bosonic new physics at any intermediate energy scale. It can be modified easily to meet the needs of the user to investigate the running of the BNV Wilson coefficients. We then compare the result with the experimental bound from the neutron-antineutron oscillation process and constrain the scale of baryon number violating new physics.

\end{abstract}

\section{Introduction} 

The observed baryon-antibaryon asymmetry in the universe remains one of the most profound mysteries in fundamental physics. With experimental evidence pushing the proton’s lifetime beyond $10^{34}$ years~\cite{Super-Kamiokande:2018apg}, simple New Physics models that violate baryon number at the TeV scale ($\mathcal{O}(1~\mathrm{TeV})$) are now strongly disfavored. However, higher-order processes that violate baryon number by two units, such as neutron-antineutron ($n{-}\bar{n}$) oscillations~\cite{Phillips:2014fgb, Super-Kamiokande:2020bov}, $nn \rightarrow \pi^0 \pi^0$~\cite{Super-Kamiokande:2015jbb}, $nn\to K^+ K^-$~\cite{Super-Kamiokande:2014hie} , deuteron decay~\cite{SNO:2017pha, Oosterhof:2019dlo}, hydrogen-antihydrogen ($H{-}\bar{H}$) oscillations~\cite{Grossman:2018rdg}, etc, remain viable due to the significantly weaker constraints on the scale of new physics. These rare processes, which break the accidental symmetries of the Standard Model (SM), serve as powerful probes for physics beyond Standard Model (BSM). A confirmed observation of any such event would have profound implications for both particle physics and cosmology. New experiments are being proposed and developed to search for these elusive phenomena~\cite{Abele:2022iml}. Nonetheless, the absence of definitive signals in terrestrial experiments continues to pose a significant challenge for theoretical models of New Physics.

Though baryon and lepton numbers are accidental symmetries of the SM at the classical level, quantum effects break them non-perturbatively \cite{tHooft:1976rip} to $U(1)_{B-L}$. But, there is no {\it a priori} reason for these symmetries to be preserved in BSM scenarios. With the proton decay suppressed, various New Physics models~\cite{Mohapatra:1980qe, Mohapatra:1980de, Mohapatra:1996pu, Pasupathy:1982qr, Rao:1982gt, MOHAPATRA19891,Babu:2008rq, Arnold:2012sd,Babu:2013yca, Berezhiani:2015afa, Dev:2015uca,Allahverdi:2017edd, Berezhiani:2020vbe, Arun:2022eqs, Thomas:2022hyj, PhysRevLett.49.7} that can accommodate baryon number violation (BNV) by two units ($\Delta B=2$) have become increasingly important, especially, with the recent observation of 11 possible candidates with an expected background of $9.3\pm2.7$ events at Super-Kamiokande~\cite{Super-Kamiokande:2020bov}, 0.37 megaton-year exposure, and the prospect of observing the neutron-antineutron oscillation at future experiments like DUNE~\cite{2817610}, Hyper-Kamiokande~\cite{Hyper-Kamiokande:2018ofw} and HIBEAM/NNBAR~\cite{Addazi:2020nlz} with much improved sensitivity. Moreover, these models can naturally host the desirable low-scale baryogenesis~\cite{Mohapatra:1986dg,Dev:2015uca,Allahverdi:2017edd,Fridell:2021gag,Gopalakrishna:2024qxk}.

Such $\Delta B=2$ processes are generated by six-quark operators~\cite{Heeck:2019kgr} of mass dimension $\geq 9$. The simplest of these are the ones that do not involve leptons, and the corresponding operators have been experimentally searched via $n-\bar{n}$ oscillations, $pp\to \pi^+\pi^+$, $nn\to \pi^0 \pi^0$, deuteron decay, and other dinucleon decay processes given in Tab.~\ref{tab:processes}. Out of these processes, $nn\to \pi^0 \pi^0$ decay and $n-\bar{n}$ oscillation constrain the baryon number violating new physics most stringently and constraint the BNV new physics to $\gtrsim 500$ TeV. Given that $n-\bar{n}$ oscillation has been of interest, and dedicated experiments to study the processes are currently under development, we investigate the renormalization group evolution (RGE) of the Wilson coefficients of dimension-9 six-quark operators and derive their bounds from the neutron-antineutron oscillation lifetime.
\begin{table}[]
    \centering
    \begin{tabular}{|c|c|} \hline
     Processes & Lifetime ($T$) $\times 10^{30}$ years \\ \hline 
         $pp\to \pi^+ \pi^+$ & 72 ~\cite{Super-Kamiokande:2015jbb}\\
         $nn\to \pi^0\pi^0$ & 404 ~\cite{Super-Kamiokande:2015jbb}\\
         $nn\to \pi^+\pi^-$ & 72~\cite{Super-Kamiokande:2015jbb} \\
         $nn\to K^+ K^-$ & 170 ~\cite{Super-Kamiokande:2014hie} \\
         $n \to \bar{n}$  & 360 ~\cite{Super-Kamiokande:2020bov}\\
         (bound neutron in ${}^{16}O$) & \\
         deuteron decay  & 11.8~\cite{SNO:2017pha} \\
\hline
    \end{tabular}
    \caption{$\Delta B=2$ processes generated by six-quark dimension-9 operators. The free $n-\bar{n}$ oscillation lifetime $\tau_{n-\bar{n}} = \sqrt{T_{n-\bar{n}}/R}$, where R is the nuclear suppression factor, is $4.7 \times 10^{8}$ sec at 90\%C.L.~\cite{Phillips:2014fgb,Super-Kamiokande:2020bov}.}
    \label{tab:processes}
\end{table}

Ignorant of the new physics responsible for the generation of $n-\bar{n}$ oscillations, it is advantageous to use the framework of effective field theory for systematic analysis, with the complete operator basis for the $\Delta B=2$ six-quark interactions found in \cite{Chang:1980ey, Kuo:1980ew, Ozer:1982qh, Rao:1982gt, Rao:1983sd,Gardner:2014cma,Babu:2015axa}. Though, the bounds discussed previously give a naive estimate on the scale associated with BNV new physics, such estimates remain incomplete because of the large disparity between the energy scales at which experimental constraints are imposed and the scale at which the new physics is matched with the dimension-9 BNV operators. Infact, the one-loop effects of SM particles, as well as, any possible new physics at intermediate scales, must be taken into account. Intermediate new physics is a compelling scenario since constraints from the Large Hadron Collider (LHC) do not exclude the presence of new particles coupling to light quarks at scales above $\sim 10~\mathrm{TeV}$. Moreover, the notion of a vast energy desert between the electroweak scale and the BNV scale $\sim 10^3~\mathrm{TeV}$ appears implausible. Should new degrees of freedom exist within the intermediate scale range $1$-$10^3~\mathrm{TeV}$ which couple to the light quark sector, they are expected to induce loop-level contributions to $\Delta B = 2$ processes, including $n-\bar{n}$ oscillations.

To connect these scales and account for such effects reliably, it is imperative to study the renormalization group evolution (RGE) of the dimension-9 six-quark operator Wilson coefficients from high scale $\sim \mathcal{O}(10^{3}~\text{TeV})$ to the low scale $\sim \mathcal{O}(10~\text{GeV})$. On the other hand, since we aim to investigate the influence of the intermediate new physics in neutron-antineutron oscillation bounds, we restrict our analysis of the running and matching of the Wilson coefficients to above the electro-weak scale. Below this scale, the QCD processes become progressively stronger and require the construction of non-perturbative renormalization of the corresponding six-quark operator matrix elements~\cite{Buchoff:2015qwa}. 

\begin{figure}[h]
    \centering
    \includegraphics[width=16cm, height=4.7cm]{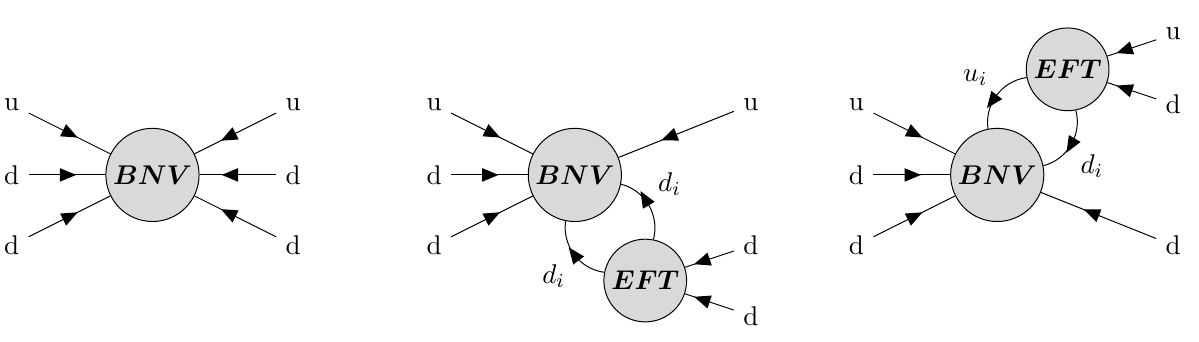}
    \caption{Intermediate scale new physics dimension-6 $(\bar{q}q)(\bar{q}q)$ effective field theory operators at entering the RG running of the dimension-9 baryon number violating operators.}
    \label{fig:1loopWC}
\end{figure}

In this work, we concentrate on the one-loop processes depicted in Fig.~\ref{fig:1loopWC}, where the operator labelled $BNV$ represents baryon number violating interactions, and the block labeled $EFT$ encapsulates the contributions from intermediate-scale new physics. These contributions generate flavor-diagonal, dimension-six four-Fermi operators of the form $(\bar{q}q)(\bar{q}q)$.

For completeness, we consider two distinct scenarios. The first assumes that the SM remains valid up to the scale at which BNV occurs, with no new physics in the intermediate regime. The second introduces new physics at an intermediate scale, responsible for generating the aforementioned dimension-six operators. In the event that future collider experiments fail to discover evidence of such intermediate-scale physics, the $EFT$ block in Fig.~\ref{fig:1loopWC} can be replaced by interactions involving only SM fields.

Although a comprehensive analysis within the complete Standard Model Effective Field Theory (SMEFT) framework is beyond the scope of this work, we aim to highlight the essential features of intermediate-scale contributions by focusing on explicit examples. Specifically, we consider two representative scenarios that give rise to the four-quark operators. The first involves additional $U(1)$ Abelian vector bosons, while the second introduces an electroweak scalar doublet, with both types of field assumed to appear at every $10~\text{TeV}$ interval up to the high scale. We compute the RGE of the baryon number violating Wilson coefficients, incorporating these intermediate-scale effects. Although the inclusion of multiple fields is in general not appealing, BSM scenarios with compactified extra dimensions, where the emergence of Kaluza Klein (KK) modes with masses $M_{KK} = n/R$ gives rise to a tower of states separated at intervals of $1/R$ (in a five-dimensional setup). Similar studies have been carried out previously, focusing on the evolution of gauge and Yukawa couplings in five and six dimensions~\cite{Bhattacharyya:2006ym,Arun:2016ela}. In particular, the RGE of coupling constants in such geometries demonstrate the occurrence of power-law running in gauge couplings, leading to unification at energies $\sim 10^3-10^6$ TeV.

To facilitate further studies, we provide a publicly available Python package~\cite{RepoName} that performs the RGE. This tool is designed to accommodate a wide class of bosonic new physics models and arbitrary choices of intermediate scales, enabling users to generalize our analysis to other scenarios. The code also includes routines for matching the resulting coefficients to current experimental bounds, enabling a consistent framework for phenomenological studies of $n-\bar{n}$ oscillations and other $\Delta B = 2$ processes. While the analysis presented here is inspired by frameworks related to extra-dimensional model building, the Python code is designed to be flexible and easily adapted to any BSM scenario the reader might wish to explore.

Although the particles within the loop in Fig.~\ref{fig:1loopWC} can be any of the SM fermions, for clarity and simplicity, we restrict them such that the minimal set of BNV operator basis given in~\cite{Ozer:1982qh} is not enlarged. While the original analysis in~\cite{Ozer:1982qh} did not incorporate Yukawa interactions, we extend the BNV operator basis to account for scalar-mediated interactions, thereby capturing additional effects relevant in realistic new physics scenarios. 

The article is organized as follows. In Sec.~\ref{sec:sixquarkoperators}, we briefly introduce the dimension-9 six-quark operators that contribute to the neutron-antineutron oscillation, and in Sec.~\ref{sec:interactions}, the SM notations and conventions used in the article. Anomalous dimensions for RGE are computed in Sec.~\ref{sec:anomalousdim}. Here, we also introduce two specific new physics scenarios: the first involving additional scalar doublets, and the second vector bosons. Both these scenarios are assumed to exist at every 10 TeV interval from the low-scale to the high-scale. In Sec.~\ref{sec:scaleChPT}, we derive the relation between the Wilson coefficients and the scale of new physics using Chiral perturbation theory, and in Sec.~\ref{sec:scalelimit}, we compute constraints on the Wilson coefficients based on the experimental bounds on the process. The python code that is used for this computation is introduced in Sec.~\ref{sec:PythonCode}. This code can be used to compute any bosonic new physics and can be generalized to include different gauge groups and new physics scales. In Sec.~\ref{sec:summary}, we summarize our results. And in Appendix.~\ref{sec:AppendixA} and Appendix.~\ref{sec:AppendixB}, we give the relevant computations and techniques used in the article.
\section{$\Delta B=2 $ dimension-9 six-quark operators} 
\label{sec:sixquarkoperators}
A strong indication of physics outside of the SM is baryon number violation. In particular, effective six-quark operators that arise at dimension-nine and are suppressed by five powers of a high energy scale \(\Lambda\), mediate neutron-antineutron (\(n-\bar{n}\)) oscillations, which involve \(\Delta B = 2\) transitions~\cite{Kuo:1980ew, Ozer:1982qh, Babu:2013yca}. Following the framework established in a series of works~\cite{Jenkins:2013zja, Jenkins:2013wua, Alonso:2013hga,Alonso:2014rga, Jenkins:2017jig, Jenkins:2017dyc}, we classify all possible \(\Delta B = 2\) six-quark operators that are invariant under the SM gauge group \(SU(3)_C \times SU(2)_L \times U(1)_Y\), since renormalizable operators within the SM cannot violate baryon number by two units. 

The effective operators for the six quarks, which change the baryon number by two units for the first generation of quarks and are invariant under $SU(3)_C \times SU(2)_L \times U(1)_Y$ without the Higgs interaction, are given as follows:
\begin{align}
    \mathcal{O}^{(1)} &= \Big[ (\bar{d}^c_{iL} u_{jL})(\bar{d}^c_{kL} u_{lL})(\bar{d}^c_{mR} d_{nR}) \notag \\
    & \quad - (\bar{d}^c_{iL} d_{jL})(\bar{u}^c_{kL} u_{lL})(\bar{d}^c_{mR} d_{nR}) \Big] \Gamma^{(1)}_{ijklmn} \label{eq:op_1_LEFT},  \\
    \mathcal{O}^{(2)} &= (\bar{d}^c_{iL} u_{jL})(\bar{d}^c_{kL} u_{lL})(\bar{d}^c_{mR} d_{nR}) \Gamma^{(2)}_{ijklmn},  \\
    \mathcal{O}^{(3)} &= (\bar{d}^c_{iL} u_{jL})(\bar{d}^c_{kR} u_{lR})(\bar{d}^c_{mR} d_{nR}) \Gamma^{(2)}_{ijklmn}, \\
    \mathcal{O}^{(4)} &= (\bar{u}^c_{iR} u_{jR})(\bar{d}^c_{kR} d_{lR})(\bar{d}^c_{mR} d_{nR}) \Gamma^{(1)}_{ijklmn},\\
    \mathcal{O}^{(5)} &= (\bar{d}^c_{iR} u_{jR})(\bar{d}^c_{kR} u_{lR})(\bar{d}^c_{mR} d_{nR}) \Gamma^{(1)}_{ijklmn}, \\
    \mathcal{O}^{(6)} &= (\bar{d}^c_{iR} u_{jR})(\bar{d}^c_{kR} u_{lR})(\bar{d}^c_{mR} d_{nR}) \Gamma^{(2)}_{ijklmn}.
\end{align}

\noindent where,

\begin{align}
    \Gamma^{(1)}_{ijklmn} &= \epsilon_{ikm} \epsilon_{jln} + \epsilon_{jkm} \epsilon_{iln} + \epsilon_{ilm} \epsilon_{jkn} \notag + \epsilon_{jlm} \epsilon_{ikn}, 
\end{align}
and
\begin{align}
    \Gamma^{(2)}_{ijklmn} &= \epsilon_{ikl} \epsilon_{jmn} + \epsilon_{jkl} \epsilon_{imn}
\end{align}

Here, $L$ and $R$ represent the left- and right-handed fields. $d_{iL}$ and $d^c_{iL} = (d_{iL})^c$ are $SU(2)_L$ left-handed doublets, while $d_{iR}$ are right-handed $SU(2)_L$ singlets. $\Gamma^{(1,2)}_{ijklmn}$ are $SU(3)_c$ tensors that ensure that the operators are $SU(3)_c$ singlets.

The Yukawa couplings and their chiral mixing effects permit three additional operators following electroweak symmetry breakdown or in the presence of new scalar interactions~\cite{Jenkins:2013wua, Jenkins:2017dyc}: 

\begin{align}
    \mathcal{O}^{(7)} &= (\bar{d}^c_{iL} u_{jL})(\bar{d}^c_{kL} u_{lL})(\bar{d}^c_{mR} d_{nR}) \Gamma^{(1)}_{ijklmn},  \\
    \mathcal{O}^{(8)} &= (\bar{d}^c_{iL} u_{jL})(\bar{d}^c_{kR} u_{lR})(\bar{d}^c_{mR} d_{nR}) \Gamma^{(1)}_{ijklmn}, \\
    \mathcal{O}^{(9)} &= (\bar{u}^c_{iR} u_{jR})(\bar{d}^c_{kR} u_{lR})(\bar{d}^c_{mR} d_{nR}) \Gamma^{(2)}_{ijklmn} \label{eq:op_9_LEFT} 
\end{align}

The Yukawa and electroweak corrections allow these other operators to mingle exclusively with quark operators, making them particularly important when scaling the theory from a high new physics scale down to the hadronic scale.

While transitioning from the LEFT to SMEFT framework, the above operators are written in a manner consistent with the unbroken $SU(2)_L \times U(1)_Y$ gauge symmetry of the SM at high energies~\cite{Alonso:2013hga, Jenkins:2017jig}. This requires writing the operators in terms of a full $SU(2)_L$-doublet $Q_L$ in the place of $d_L$ and $u_L$.

In this approach, left-handed quark fields are promoted to full electroweak doublets:
\[
Q_L = \begin{pmatrix} u_L \\ d_L \end{pmatrix},
\quad \phi = \begin{pmatrix} \phi^+ \\ \phi^0 \end{pmatrix},
\]

and the operators are rewritten accordingly to maintain full \(SU(2)_L \times U(1)_Y\) invariance.

For example, operator \(\mathcal{O}^{(1)}\) becomes:
\begin{align}
    \mathcal{O}^{(1)}_\text{SMEFT} &= \Big[ (\bar{Q}^c_{iL} Q_{jL})(\bar{Q}^c_{kL} Q_{lL})(\bar{d}^c_{mR} d_{nR}) \notag \\
    & \quad - (\bar{Q}^c_{iL} Q_{jL})(\bar{Q}^c_{kL} Q_{lL})(\bar{d}^c_{mR} d_{nR}) \Big] \Gamma^{(1)}_{ijklmn}.
\end{align}

This embedding guarantees gauge invariance is maintained in the high-energy limit and permits consistent RGE across scales~\cite{Alonso:2014rga}.

\section{Interactions}
\label{sec:interactions}
\subsection{Standard Model}
The whole set of SM interactions, especially those involving the Higgs field and gauge bosons, must be taken into account in order to properly understand the evolution of the renormalization group and the mixing of baryon number violating six-quark operators. These interactions drive the evolution of effective operators between energy scales and directly contribute to operator mixing at one loop~\cite{Mohapatra1980,Rao:1983sd, Alonso_2014, PhysRevD.100.015032}.

The Lagrangian, including the gauge terms and the Higgs interaction for the first-generation quarks, is given by:
\begin{align}
\mathcal{L} &= \mathcal{L}_{gauge} + \mathcal{L}_{Yukawa} \\
&= \, g_3 \left[\overline{u}_L \gamma^\mu \frac{\lambda_a}{2} u_L + \overline{d}_L \gamma^\mu \frac{\lambda_a}{2} u_L + (\textit{L $\to$ R}) \right] G_\mu^a \notag\\
& + g_2 \sum_{i=r,g,b} \left( \overline{u}_{iL}, \overline{d}_{iL} \right) \overrightarrow{W}_\mu \frac{\vec{\tau}}{2} \gamma^\mu 
\begin{pmatrix}
u_{iL} \\ 
d_{iL}
\end{pmatrix} \notag\\
& + g_1 \sum_{i=r,g,b} \left[ \frac{1}{6} \overline{u}_{iL} \gamma^\mu u_{iL} + \frac{1}{6} \overline{d}_{iL} \gamma^\mu d_{iL} + \frac{2}{3} \overline{u}_{iR} \gamma^\mu u_{iR} - \frac{1}{3} \overline{d}_{iR} \gamma^\mu d_{iR} \right] B_\mu^0 \notag\\
& - \left[ y_u \, \overline{Q}_L \tilde{\phi} u_R + y_d \, \overline{Q}_L \phi d_R + \text{h.c.} \right],
\end{align}
where,
\begin{itemize}
    \item $Q_L = \begin{pmatrix} u_L \\ d_L \end{pmatrix}$ is the left-handed quark doublet,
    \item $\phi = \begin{pmatrix} \phi^+ \\ \phi^0 \end{pmatrix}$ is the Higgs doublet,
    \item $\tilde{\phi} = i\sigma^2 \phi^*$ is the hypercharge-conjugate Higgs,
    \item $\lambda_a$ and $\tau^j$ are the SU(3)$_C$ and SU(2)$_L$ generators,
    \item $g_1$, $g_2$, $g_3$ are the gauge couplings of $U(1)_Y$, $SU(2)_L$, and $SU(3)_C$ respectively.
\end{itemize}

Here, $a$  = 1,2, ..., 8 and the $\lambda_a$ are the SU(3)$_C$ matrices. $\tau^j$ ($j$ = 1,2,3) are Pauli matrices. $r$, $g$, and $b$ are the abbreviations for red, green, and blue colour charges; the couplings of the $B^0$ gauge boson of U(1)$_Y$ are read off from the expression for the electric charge 
\begin{align}
    Q = T^3_L + Y
\end{align}
After electroweak symmetry breaking, the Higgs field acquires a vacuum expectation value (VEV)
\begin{equation}
\phi = \frac{1}{\sqrt{2}} \begin{pmatrix} 0 \\ v + h \end{pmatrix}, \quad \text{with } v \approx 246\ \text{GeV},
\end{equation}
leading to fermion masses through the Yukawa terms
\begin{equation}
\mathcal{L}_{\text{Higgs}} = -\left( \frac{y_u v}{\sqrt{2}} \bar{u}_L u_R + \frac{y_d v}{\sqrt{2}} \bar{d}_L d_R + \frac{y_u}{\sqrt{2}} \bar{u}_L u_R h + \frac{y_d}{\sqrt{2}} \bar{d}_L d_R h + \text{h.c.} \right),
\end{equation}
where the quark masses are given by,
\begin{equation}
m_u = \frac{y_u v}{\sqrt{2}}, \quad m_d = \frac{y_d v}{\sqrt{2}}.
\end{equation}
Similarly, the $W$ bosons acquire mass via the Higgs kinetic term
\begin{equation}
\mathcal{L}_{W^\pm} = \frac{1}{2} g_2^2 \frac{v^2}{2} W^+_\mu W^{-\,\mu}, \quad M_W = \frac{g_2 v}{2}.
\end{equation}

\subsection{Additional $U(1)_Y$ vector boson new physics}

To explore the RGE of baryon number violating operators in the presence of extended gauge sectors, we consider a minimal extension of the SM in which $J$ additional $U(1)$-like gauge interactions are introduced. These new $U(1)_Y$-like groups are denoted as $U(1)_{Y_j}$, with $j = 1, 2, \dots, J$. Each of these new Abelian symmetries couples to the quark fields with its own independent gauge coupling constant, denoted by $g_{1j}$.

To distinguish this extended setup from the SM, the original $U(1)_Y$ gauge coupling is relabeled as $g_{10}$, and the corresponding gauge boson is now denoted as $B_\mu^{(0)}$. The additional $J$ gauge bosons corresponding to the new $U(1)_{Y_j}$ groups are denoted by $B_\mu^{(j)}$, where $j = 1, 2, \dots, J$.

The modified gauge interaction part of the Lagrangian, focusing on the Abelian sectors, becomes
\begin{align}
\mathcal{L}_{\text{Abelian}}^{\text{extended}} = \sum_{j=0}^{J} g_{1j} \sum_{i=r,g,b} \Bigg[ &\frac{1}{6} \overline{u}_{iL} \gamma^\mu u_{iL} + \frac{1}{6} \overline{d}_{iL} \gamma^\mu d_{iL} + \frac{2}{3} \overline{u}_{iR} \gamma^\mu u_{iR} - \frac{1}{3} \overline{d}_{iR} \gamma^\mu d_{iR} \Bigg] B_\mu^{(j)},
\end{align}
where,
\begin{itemize}
    \item $g_{1j}$ is the gauge coupling associated with the $j^{\text{th}}$ $U(1)_{Y_j}$ gauge group,
    \item $B_\mu^{(j)}$ is the corresponding gauge boson, with $B_\mu^{(0)}$ representing the SM hypercharge boson,
    \item The quark hypercharges are taken to be identical under all $U(1)_{Y_j}$ groups for simplicity, though generalization to non-universal charge assignments is possible.
\end{itemize}

This modification sets the stage for computing the RGE of effective operators under the influence of an enlarged Abelian gauge structure, with operator mixing now potentially occurring through all $J+1$ $U(1)_Y$ gauge interactions.

\subsection{Additional scalar doublet boson new physics}

In continuity with the previous subsection, we consider a generalization of the SM in which $N$ additional scalar fields are introduced. These fields, denoted by $\phi_n$ for $n = 1, 2, \dots, N$, are electroweak doublets transforming identically to the SM Higgs doublet under $SU(2)_L \times U(1)_Y$.

The SM Yukawa couplings $y_u$ and $y_d$ are now redefined as $y_{u0}$ and $y_{d0}$ to label the couplings associated with the original Higgs doublet $\phi_0$. Each new scalar doublet $\phi_n$ is associated with its up- and down-type Yukawa couplings, $y_{un}$ and $y_{dn}$, respectively.

The extended Yukawa sector of the Lagrangian becomes:
\begin{align}
\mathcal{L}_{\text{Yukawa}}^{\text{extended}} = -\sum_{n=0}^{N} \left[ y_{un} \, \overline{Q}_L \, \tilde{\phi}_n \, u_R + y_{dn} \, \overline{Q}_L \, \phi_n \, d_R + \text{h.c.} \right],
\end{align}
where,
\begin{itemize}
    \item $\phi_n = \begin{pmatrix} \phi_n^+ \\ \phi_n^0 \end{pmatrix}$ is the $n^\text{th}$ Higgs-like $SU(2)_L$ doublet,
    \item $\tilde{\phi}_n = i \sigma^2 \phi_n^*$ is the hypercharge-conjugate of $\phi_n$,
    \item $y_{un}$ and $y_{dn}$ are the Yukawa couplings for up- and down-type quarks for the $n^\text{th}$ scalar doublet.
\end{itemize}

\begin{figure}[h!]
    \centering
    \includegraphics[scale=0.15]{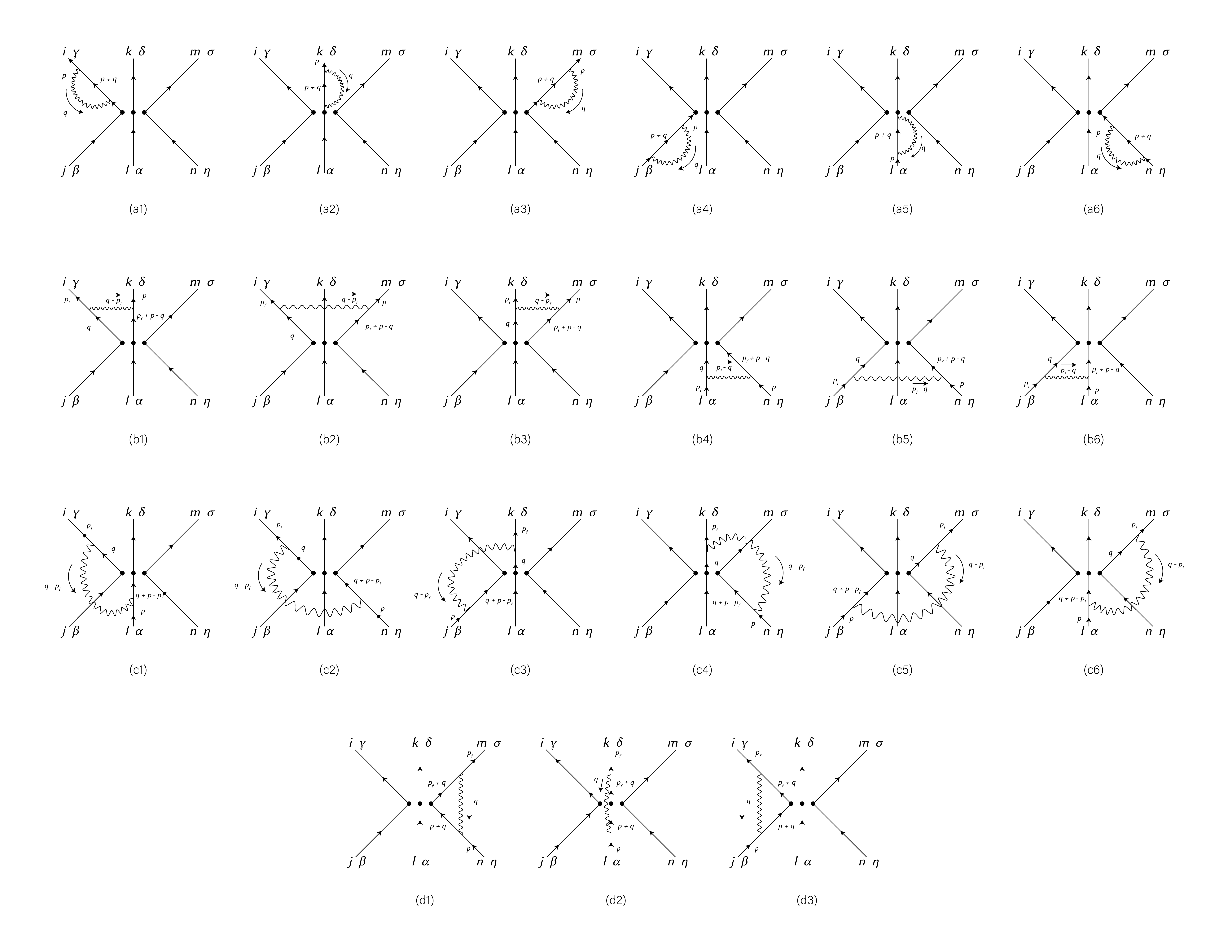}
    \caption{All possible 1-boson-exchange diagrams for $3q\rightarrow3\bar{q}$ processes. }
    \label{fig:Poss_21_diag}
\end{figure}

The anomalous dimension ($\gamma^{(n)}_i$) of the six-quark operators coming from the low energy SU(3)$_C$, SU(2)$_L$, and U(1)$_Y$ subgroups can be calculated by examining all the possible 21 diagrams, which are segmented into 4 sets of graphs depending on their Dirac algebra in Fig.~\ref{fig:Poss_21_diag}. The explicit calculation to find the anomalous dimension contribution for $W$-Boson and Higgs interaction in the \textit{Landau Gauge} is shown in the Appendix.~\ref{sec:AppendixA} and Appendix.~\ref{sec:AppendixB} for a specific case of operator $\mathcal{O}^{(2)}$. The full anomalous dimension matrix consisting of gluon, $W$-boson, $B^0$-boson, and Higgs interaction is calculated in the following section. 

\section{Anomalous dimensions}
\label{sec:anomalousdim}
In this section, we calculate the anomalous dimensions of six-quark operators in the presence of SM and new physics interactions. The evolution of the Wilson coefficients \( \widetilde{C}^{(n)}(\mu) \) of effective operators, in the background of the SM fields, via the Callan-Symanzik equation is given by,
\begin{align}
    \left[\mu \frac{\partial}{\partial \mu} + \sum_{k=1,2,3} \beta_k \frac{\partial}{\partial g_k} + \sum_{k'=u,d} \beta_{y_{k'}} \frac{\partial}{\partial y_{k'}} + \gamma^{(n)} \right] \times \widetilde{C}^{(n)}  = 0,
    \label{eq:CS}
\end{align}

where, the beta functions of the gauge couplings \( g_1, g_2, g_3 \) associated with the \( U(1)_Y \), \( SU(2)_L \), and \( SU(3)_C \) gauge groups are represented by \( \beta_1, \beta_2, \beta_3 \) and the beta functions of the up- and down-type Yukawa couplings are indicated by \( \beta_{y_{u,d}} \). The scale dependence of the six-quark operator basis is captured by the anomalous dimension matrix \( \gamma^{(n)} \), which is calculated at one-loop order using the \( \overline{\text{MS}} \) method and dimensional regularization.

These beta functions are expressed as,
\begin{align}
    \beta_k = - \frac{g_k^3}{16 \pi^2} \beta_k^0 \quad (k = 1, 2, 3), \quad \beta_{y_{k'}} = - \frac{y_{k'}}{16 \pi^2} \beta_{y_{k'}}^0 \quad (k' = u,d)
\end{align}
where,
\begin{align}
    \beta_1^0 &= - \frac{2}{3} N_f - \frac{1}{10}, \quad
    \beta_2^0 = \frac{22}{3}- \frac{2}{3} N_f - \frac{1}{6},\quad
    \beta_3^0 = 11 - \frac{2}{3} N_f, \notag \\
    \beta_{y_u} &= \frac{3}{2} (y^2_u - y^2_d) + T - \left(\frac{17}{20} g_1^2 + \frac{9}{4} g_2^2 + 8 g_3^2 \right), \notag \\
    \beta_{y_d} &= \frac{3}{2} (y^2_d - y^2_u) + T - \left(\frac{1}{4} g_1^2 + \frac{9}{4} g_2^2 + 8 g_3^2 \right) \label{eq:bfn_sm}
\end{align}
with \( T = 3(y_u^2 + y_d^2) \) and \( N_f = 6 \) representing the  number of quark flavors ~\cite{Buchalla:1995vs,Manohar:1994kq}.

The RG equations define the anomalous dimension matrices,
\begin{align}
    \dot{\widetilde{C}}^{(n)}i(\mu) &= 16 \pi^2 \mu \frac{d}{d\mu} {\widetilde{C}}^{(n)}i(\mu) = (\gamma_{\tilde{C}}^{(n)})_{ij} {\widetilde{C}}^{(n)}j(\mu),\label{eq:WC_An_dim}
\end{align}
where, $\gamma_C = \gamma^T$ with $\gamma$ given by the matrix $\hat{\mathcal{O}}$ as,
\begin{align}
    \frac{\gamma}{16\pi^2} &= \hat{\mathcal{O}}^{-1} \frac{d \hat{\mathcal{O}}}{d \ln \mu}.
\end{align}

The Callan-Symanzik equation, in Eq.~\ref{eq:CS}, is generalized easily to include the interactions of bosonic new physics by computing the beta functions and the anomalous dimensions in the new background. 

For clarity and simplicity, here, we take a specific case of the operator $\mathcal{O}^{(2)}$ and calculate its anomalous dimension, first with the SM background and then with the new physics background. The calculations for the rest of the operators are straightforward and can be generalized from this. Not all of the diagrams given in Fig.~\ref{fig:Poss_21_diag} contribute to the anomalous dimensions of all the operators. A complete list is shown in the table .~\ref{tab:1loopdiagramsops}.

\begin{table}[]
    \centering
    \begin{tabular}{c|c}
      Operators   &  1-loop diagrams \\[1ex] \hline \\
    $  \mathcal{O}^{(1)} $ &(a1)--(a6), (c1), (c3), and (d1)--(d3)  \\
    $  \mathcal{O}^{(2)} $ &   (a1)--(a6), (c1), (c3), and (d1)--(d3)\\
    $  \mathcal{O}^{(3)} $ &(a1)--(a6), (c6), and (d1)--(d3)  \\
    $  \mathcal{O}^{(4)} $ &(a1)--(a6), (c1)--(c3), (c5), and (d1)--(d3)  \\
    $  \mathcal{O}^{(5)} $ &(a1)--(a6), (c1), (c3), (c5), (c6), and (d1)--(d3)  \\
    $  \mathcal{O}^{(6)} $ &(a1)--(a6), (c3), (c5), (c6), and (d1)--(d3)  \\
    $  \mathcal{O}^{(7)} $ &(a1)--(a6), (c1), (c3), and (d1)--(d3)  \\     
    $  \mathcal{O}^{(8)} $ &(a1)--(a6), (c6), and (d1)--(d3)  \\     
    $  \mathcal{O}^{(9)} $ &(a1)--(a6), (c1)--(c3), (c5), and (d1)--(d3)  \\
    \end{tabular}
    \caption{The 1-loop diagrams that contribute to the anomalous dimension calculation of the operators among all the diagrams in Fig.~\ref{fig:Poss_21_diag}}
    \label{tab:1loopdiagramsops}
\end{table}

\subsection{Calculation of anomalous dimension in the presence of Standard Model fields}

From the 21 diagrams contributing to the one-loop correction of $\mathcal{O}^{(2)}$, only diagrams (a1)--(a6), (c1), (c3), (d2), and (d3) involve the Yukawa interaction. Among these, diagrams (d1)--(d3) arise from interactions mediated by the gluon, \( W_L \), and \( B^0 \) bosons, under the \textit{Landau gauge}.

\subsubsection{Self-Energy Correction}

Diagrams (a1)--(a6) correspond to fermion self-energy corrections and yield identical counterterms. For clarity and simplicity, we explicitly present the calculation for diagram (a3). The corresponding diagram is shown in Fig.~\ref{fig:Ap_A_Yukawa_self_One_loop}, and the full computation is provided in Appendix.~\ref{sec:AppendixA}.

The renormalized 1PI two-point function at one-loop order for Fig.~\ref{fig:Ap_A_Yukawa_self_One_loop} is:
\begin{align}
    i \Gamma(\not{p}) = i(\not{p} - M) - i \Sigma(\not{p}) + i (\delta Z^{\phi}_{\mathcal{O}^{(2)}(a3)} \not{p} - \delta M) \ ,
\end{align}

where \( \Sigma(\not{p}) \) is the loop correction, and \( \delta Z^{\phi}_{\mathcal{O}^{(2)}(a3)}, \delta M \) are the wavefunction and mass counterterms, respectively. Evaluating at zero momentum, we impose:
\begin{align}
    i \Sigma(0) + \delta Z^{\phi}_{M(a3)} = 0 \\
    {\frac{d}{d\not{p}}\Sigma(\not{p})}\Bigr\rvert_{\not{p} = 0} - \delta Z^{\phi}_{\mathcal{O}^{(2)}(a3)} = 0
\end{align}

Neglecting the mass counterterms, we extract the wavefunction renormalization term using dimensional regularization with Higgs mass \( m \) and fermion mass \( M \).

\begin{align}
    - i \Sigma(\not{p}) &= \mu^{\epsilon}{y_d}^2 \int \frac{d^d q}{(2\pi)^d} \frac{i}{q^2 - m^2}  \frac{i(\not{q} + \not{p} - M)}{(q + p)^2 - M^2}\\
    &= {y_d}^2 \frac{1}{(4\pi)^2} \frac{1}{\epsilon} \frac{1}{2} + \text{finite}
\end{align}

Hence, the corresponding wavefunction counterterm for all three color contractions is:
\begin{align}
    \delta Z^{\phi}_{\mathcal{O}^{(2)}(a3)} = -{y_d}^2 \frac{3}{(4\pi)^2} \frac{1}{\epsilon} \frac{1}{2} + \text{finite}.
\end{align}

\subsubsection{Vertex Correction}

The six-point vertex corrections arising from Yukawa interactions are represented by diagrams (c2), (c3), (d2), and (d3). For clarity and simplicity, we focus on the calculation for diagram (d3), which involves Higgs exchange and captures the dominant contribution to the Yukawa-induced one-loop corrections. The corresponding diagram is shown in Fig.~\ref{fig:Ap_A_Yukawa_vertex_One_loop}, and the complete computation is provided in Appendix.~\ref{sec:AppendixA}.

Yukawa interactions can induce mixing between operators of different structures through loop corrections. Specifically, an operator of the form \( \mathcal{O}^{(3)} \) can generate contributions to the renormalization of \( \mathcal{O}^{(2)} \), typically a purely fermionic operator, through vertex-type diagrams. This mixing occurs because loop diagrams involving the Yukawa vertex lead to non-zero off-diagonal entries in the anomalous dimension matrix. Consequently, even if \( \mathcal{O}^{(3)} \) is not present at tree level, it can effectively shift the Wilson coefficient of \( \mathcal{O}^{(2)} \) during renormalization group running, encoding the physical impact of Yukawa couplings at higher loops.

In Fig.~\ref{fig:Ap_A_Yukawa_vertex_One_loop}, we denote the incoming and outgoing fermion momenta as \( p_i \) and \( p_f \), respectively. For Yukawa interactions, the renormalized 1PI six-point function at one-loop order is given by:
\begin{align}
-i\Gamma(p_f, p_i) &= \widetilde{C}^{(2)} - iV(p_f, p_i) + \delta Z^{\phi}_{\mathcal{O}^{(2)}(d3)}
\end{align}

at zero momentum,
\begin{align}
-i\Gamma(0, 0) &= \widetilde{C}^{(2)} - iV(0, 0) + \delta Z^{\phi}_{\mathcal{O}^{(2)}(d3)} = \widetilde{C}^{(2)}\\
\delta Z^{\phi}_{\mathcal{O}^{(2)}(d3)} &= -iV(0, 0)
\end{align}

The counter term is calculated using the following diagram: the mass of Higgs is calculated as $m$ and the mass of fermion as $M$ using dimensional regularization.
\begin{align}
-iV(p_f, p_i) &= -\widetilde{C}^{(3)}{y_d}{y_u} \int \frac{d^dq}{(2\pi)^d} \frac{i}{q^2 - m^2} \frac{i}{\not{p}_f + \not{q} - M} \frac{i}{\not{p}_i + \not{q} - M}
\end{align}

Hence, the corresponding wavefunction counterterms for all three color contractions are:
\begin{align}
    \delta Z^{\phi}_{\mathcal{O}^{(2)}(d3)}  = \widetilde{C}^{(3)}{y_d}{y_u} \frac{3}{(4\pi)^2} \frac{1}{\epsilon} + \text{finite}.
\end{align}

Combining the contribution from diagrams (a1)--(a6), (c1), (c3), (d2), and (d3):
\begin{align}
    \delta Z^{\phi}_{\mathcal{O}^{(2)}}  = \Bigg[\frac{12\widetilde{C}^{(3)} y_u y_d}{(4\pi)^2} + \frac{3\widetilde{C}^{(2)}( (y_u)^2 + 2(y_d)^2)}{2(4\pi)^2}\Bigg] \frac{1}{\epsilon} + \text{finite}.
\end{align}

In the case of gauge boson interactions, the diagrams (d1)--(d3) correspond to the gluon, \( W_L \), and \( B^0 \) bosons, respectively, under the \textit{Landau Gauge}. To streamline the discussion, we compute only the contribution from diagram (d1) as shown in Fig.~\ref{fig:Ap_B_W_L_Vertex_One_loop}, which provides a representative example for each of the three gauge sectors with gluons and $B^0$-boson are treated as massless. The full computation is provided in Appendix.~\ref{sec:AppendixB}.

In Fig.~\ref{fig:Ap_B_W_L_Vertex_One_loop}, we denote the incoming and outgoing fermion momenta as \( p_i \) and \( p_f \), respectively. For \( W_L \) bosons interactions, the renormalized 1PI six-point function at one-loop order is given by:
\begin{align}
\Gamma_{ijklm'n'}^{(2)} \Bigg[\frac{\lambda_a}{2}\Bigg]_{mm'} \Bigg[\frac{\lambda_a}{2}\Bigg]_{nn'} i \widetilde{C}^{(2)} \mu^{\epsilon/2} \Lambda^{\text{loop}}(p_f, p_i)
\end{align}

where,
\begin{align}
 \Lambda^{\text{loop}}(p_f, p_i) &= 2(i g_1 \mu^{\epsilon/2})^2 \int \frac{d^dq}{(2\pi)^d} \Bigg[\frac{g_{\rho\sigma}}{q^2 - M^2 + i\epsilon} \nonumber \gamma^\rho \frac{i (\not{q} + \not{p_f} + m')}{(p_f + q)^2 - m'^2 + i\epsilon} \frac{-i (\not{q} + \not{p_i}  + m)}{(p_i + q)^2 - m^2 + i\epsilon} \gamma^\sigma\\ 
&- \frac{((1-\xi)i)q_\rho q_\sigma}{[q^2 - M^2 + i\epsilon]^2} \gamma^\rho \frac{i (\not{q} + \not{p_f} + m')}{(p_f + q)^2 - m'^2 + i\epsilon} \frac{-i (\not{q} + \not{p_i}  + m)}{(p_i + q)^2 - m^2 + i\epsilon} \gamma^\sigma\Bigg]
\end{align}

We decompose the vertex correction into a gauge-invariant and a gauge-dependent part:
\begin{align}
    \Lambda^{\text{loop}}(p_f, p_i) = \Lambda_A^{\text{loop}}(p_f, p_i) -(1-\xi)\Lambda_B^{\text{loop}}(p_f, p_i)\label{eq:six_1_loop_1}
\end{align}

\noindent where $\Lambda^{\text{loop}}$ is related to the full vertex $\Gamma$ through the relation
\begin{align}
i \Gamma(p_f,p_i) &= i({\widetilde{C}}^{(2)} + \Lambda^{\text{loop}}(p_f, p_i) + \delta Z_{\mathcal{O}^{(2)}})
\end{align}

Therefore, the counter term for the Landau gauge (for which the gauge parameter
${\xi}$ = 0) will be,
\begin{align}
\delta Z^{W_L}_{\mathcal{O}^{(2)}(d1)} &= \Gamma_{ijklm'n'}^{(2)} \Bigg[\frac{\lambda_a}{2}\Bigg]_{mm'} \Bigg[\frac{\lambda_a}{2}\Bigg]_{nn'} \delta Z'^{W_L}_{\mathcal{O}^{(2)}}
\end{align}

where,
\begin{align}
\delta Z'^{W_L}_{\mathcal{O}^{(2)}} &= - 2(i \widetilde{C}^{(2)} \mu^{\epsilon/2})\Lambda^{\text{loop}}(p_f, p_i) \Big|_{\not{p}_i,\not{p}_f=0}\\
& = 2\widetilde{C}^{(2)}g_1^2 \frac{(3)}{(4\pi)^2} \frac{1}{\epsilon} + \text{finite}
\end{align}

With this, the counterterm for all three colour contractions can be written as, 
\begin{align}
\delta Z^{W_L}_{\mathcal{O}^{(2)}(d1)} &= \Gamma_{ijklm'n'}^{(2)} \Bigg[\frac{\lambda_a}{2}\Bigg]_{mm'} \Bigg[\frac{\lambda_a}{2}\Bigg]_{nn'}  2\widetilde{C}^{(2)}g_1^2 \frac{(3)}{(4\pi)^2} \frac{1}{\epsilon} + \text{finite}
\end{align}

We use the $SU(n)$ relation to obtain~\cite{Nishi2005},
\begin{align}
\delta Z^{W_L}_{\mathcal{O}^{(2)}(d1)} &= \frac{1}{2}\Bigg[\delta_{mn'}\delta_{nm'}-\frac{1}{2}\delta_{mm'}\delta_{nn'}\Bigg]  2\widetilde{C}^{(2)}g_1^2 \frac{(3)}{(4\pi)^2} \frac{1}{\epsilon} + \text{finite}\\
&=  \frac{1}{2}\Bigg[\frac{3}{2}\Bigg]  2\widetilde{C}^{(2)}g_1^2 \frac{(3)}{(4\pi)^2} \frac{1}{\epsilon} + \text{finite}\\
&=  \frac{9}{2}\widetilde{C}^{(2)}g_1^2 \frac{1}{(4\pi)^2} \frac{1}{\epsilon} + \text{finite}
\end{align}

Combining the contribution from diagrams d1, d2, and d3:
\begin{align}
\delta Z^{W_L}_{\mathcal{O}^{(2)}} &=  9\widetilde{C}^{(2)} \frac{g_1^2}{(4\pi)^2} \frac{1}{\epsilon} + \text{finite}
\end{align}

Similarly we can write counterterms coefficient for gluons and $B^0$-bosons for $\mathcal{O}^{(2)}$ as in Tab.~\ref{table:O2counterterms},

\begin{table}[h!]
    \centering
    \begin{tabular}{ccccc}
        \toprule
        \textbf{Interaction} \\ \textbf{counter term} & $Higgs$ & $W_L$ & $gluon$ & $B^0$ \\
        \midrule
        $\delta Z_{\mathcal{O}^{(2)}}$ & $\frac{12\widetilde{C}^{(3)} y_u y_d}{(4\pi)^2} + \frac{3\widetilde{C}^{(2)}( (y_u)^2 + 2(y_d)^2)}{2(4\pi)^2}$ & $\widetilde{C}^{(2)} \frac{9g_2^2}{(4\pi)^2}$ & $\widetilde{C}^{(2)} \frac{8g_3^2}{(4\pi)^2}$ & $\widetilde{C}^{(2)} \frac{-g_1^2}{(4\pi)^2}$  \\
        \bottomrule
    \end{tabular}
    \caption{Counterterm coefficients for operator $\mathcal{O}^{(2)}$ from interactions with Higgs, $W_L$, gluon, and $B^0$ gauge bosons.}
    \label{table:O2counterterms}
\end{table}

We can write counterterms coefficient for all nine operators as given in Tab.~\ref{table:allCounterterms},

\begin{table}[h!]
    \centering
    \begin{tabular}{ccccc}
        \toprule
        \textbf{Interaction} \\ \textbf{counter term} & $Higgs$ & $W_L$ & $gluon$ & $B^0$ \\
        \midrule
        $\delta Z_{\mathcal{O}^{(1)}}$ & $\frac{6\widetilde{C}^{(7)} y_u y_d}{(4\pi)^2}$ & $0$ & $0$ & $\widetilde{C}^{(1)} \frac{-g_1^2}{(4\pi)^2}$  \\
        
        $\delta Z_{\mathcal{O}^{(2)}}$ & $\frac{12\widetilde{C}^{(3)} y_u y_d}{(4\pi)^2} + \frac{3\widetilde{C}^{(2)}( (y_u)^2 + 2(y_d)^2)}{2(4\pi)^2}$ & $\widetilde{C}^{(2)} \frac{9g_2^2}{(4\pi)^2}$ & $\widetilde{C}^{(2)} \frac{8g_3^2}{(4\pi)^2}$ & $\widetilde{C}^{(2)} \frac{-g_1^2}{(4\pi)^2}$  \\
        
        $\delta Z_{\mathcal{O}^{(3)}}$ & $\frac{6\widetilde{C}^{(2)} y_u y_d}{(4\pi)^2} + \frac{3\widetilde{C}^{(9)} y_u y_d}{(4\pi)^2} + \frac{3\widetilde{C}^{(3)}( (y_u)^2 + 2(y_d)^2)}{2(4\pi)^2}$ & $\widetilde{C}^{(3)} \frac{9g_2^2}{2(4\pi)^2}$ & $\widetilde{C}^{(3)} \frac{8g_3^2}{(4\pi)^2}$ & $\widetilde{C}^{(3)} \frac{g_1^2}{2(4\pi)^2}$  \\
        
        $\delta Z_{\mathcal{O}^{(4)}}$ & $\frac{18\widetilde{C}^{(8)} y_u y_d}{(4\pi)^2} + \frac{3\widetilde{C}^{(4)}( (y_u)^2 + 2(y_d)^2)}{2(4\pi)^2}$ & $0$ & $0$ & $\widetilde{C}^{(4)} \frac{-4g_1^2}{(4\pi)^2}$  \\

        $\delta Z_{\mathcal{O}^{(5)}}$ & $\frac{12\widetilde{C}^{(8)} y_u y_d}{(4\pi)^2} + \frac{3\widetilde{C}^{(5)}( (y_u)^2 + 2(y_d)^2)}{2(4\pi)^2}$ & $0$ & $0$ & $\widetilde{C}^{(5)} \frac{2g_1^2}{(4\pi)^2}$  \\

        $\delta Z_{\mathcal{O}^{(6)}}$ & $\frac{18\widetilde{C}^{(3)} y_u y_d}{(4\pi)^2} + \frac{3\widetilde{C}^{(6)}( (y_u)^2 + 2(y_d)^2)}{2(4\pi)^2}$ & $0$ & $\widetilde{C}^{(6)} \frac{8g_3^2}{(4\pi)^2}$ & $\widetilde{C}^{(6)} \frac{2g_1^2}{(4\pi)^2}$  \\
        
        $\delta Z_{\mathcal{O}^{(7)}}$ & $\frac{12\widetilde{C}^{(8)} y_u y_d}{(4\pi)^2} + \frac{3\widetilde{C}^{(7)}( (y_u)^2 + 2(y_d)^2)}{2(4\pi)^2}$ & $\widetilde{C}^{(7)} \frac{9g_2^2}{(4\pi)^2}$ & 0 & $\widetilde{C}^{(7)} \frac{-g_1^2}{(4\pi)^2}$  \\

        $\delta Z_{\mathcal{O}^{(8)}}$ & $\frac{6\widetilde{C}^{(7)} y_u y_d}{(4\pi)^2} + \frac{3\widetilde{C}^{(4)} y_u y_d}{(4\pi)^2} + \frac{3\widetilde{C}^{(8)}( (y_u)^2 + 2(y_d)^2)}{2(4\pi)^2}$ & $\widetilde{C}^{(8)} \frac{9g_2^2}{2(4\pi)^2}$ & 0 & $\widetilde{C}^{(8)} \frac{g_1^2}{2(4\pi)^2}$  \\

        $\delta Z_{\mathcal{O}^{(9)}}$ & $\frac{18\widetilde{C}^{(3)} y_u y_d}{(4\pi)^2} + \frac{3\widetilde{C}^{(9)}( (y_u)^2 + 2(y_d)^2)}{2(4\pi)^2}$ & $0$ & $\widetilde{C}^{(9)} \frac{8g_3^2}{(4\pi)^2}$ & $\widetilde{C}^{(9)} \frac{-4g_1^2}{(4\pi)^2}$  \\
        \bottomrule
    \end{tabular}
    \caption{Counterterm coefficients $\delta Z_{\mathcal{O}^{(i)}}$ for all nine operators due to their interactions with Higgs, $W_L$, gluon, and $B^0$ fields.}
    \label{table:allCounterterms}
\end{table}

Hence, the Anomalous dimension matrix for all 9 operators is calculated with mixing terms as given in Eq.~\ref{eq:An_dim_mat}.
\begin{align}
    \begin{pmatrix}
        \dot{\mathcal{O}}^{(1)}(\mu) \\[6pt]
        \dot{\mathcal{O}}^{(2)}(\mu) \\[6pt]
        \dot{\mathcal{O}}^{(3)}(\mu) \\[6pt]
        \dot{\mathcal{O}}^{(4)}(\mu) \\[6pt]
        \dot{\mathcal{O}}^{(5)}(\mu) \\[6pt]
        \dot{\mathcal{O}}^{(6)}(\mu) \\[6pt]
        \dot{\mathcal{O}}^{(7)}(\mu) \\[6pt]
        \dot{\mathcal{O}}^{(8)}(\mu) \\[6pt]
        \dot{\mathcal{O}}^{(9)}(\mu)
    \end{pmatrix}
    &= 
    \begin{pmatrix}
        \gamma_{11} & 0 & 0 & 0 & 0 & 0 & \gamma_{17} & 0 & 0 \\[6pt]
        0 &  \gamma_{22} & \gamma_{23} & 0 & 0 & 0 & 0 & 0 & 0 \\[6pt]
        0 & \gamma_{32} &  \gamma_{33} & 0 & 0 & 0 & 0 & 0 & \gamma_{39} \\[6pt]
        0 & 0 & 0 & \gamma_{44} & 0 & 0 & 0 & \gamma_{48} & 0 \\[6pt]
        0 & 0 & 0 & 0 & \gamma_{55} & 0 & 0  &\gamma_{58}  & 0 \\[6pt]
        0 & 0 & \gamma_{63} & 0 & 0 & \gamma_{66} & 0 & 0 & 0 \\[6pt]
        0 & 0 & 0 & 0 & 0 & 0 & \gamma_{77} & \gamma_{78} & 0 \\[6pt]
        0 & 0 & 0 & \gamma_{84} & 0 & 0 & \gamma_{87} &  \gamma_{88} & 0 \\[6pt]
        0 & 0 & \gamma_{93} & 0 & 0 & 0 & 0 & 0 & \gamma_{99}
    \end{pmatrix}
    \begin{pmatrix}
        {\mathcal{O}}^{(1)}(\mu) \\[6pt]
        {\mathcal{O}}^{(2)}(\mu) \\[6pt]
        {\mathcal{O}}^{(3)}(\mu) \\[6pt]
        {\mathcal{O}}^{(4)}(\mu) \\[6pt]
        {\mathcal{O}}^{(5)}(\mu) \\[6pt]
        {\mathcal{O}}^{(6)}(\mu) \\[6pt]
        {\mathcal{O}}^{(7)}(\mu) \\[6pt]
        {\mathcal{O}}^{(8)}(\mu) \\[6pt]
        {\mathcal{O}}^{(9)}(\mu)
    \end{pmatrix}
    \label{eq:An_dim_mat}
\end{align}

where the matrix elements are given in Tab.~\ref{table:AnomalousMatrix}
\begin{table}[h]
    \centering
    \begin{tabular}{cccc}
        \toprule
        $\boldsymbol{\gamma_{ij}}$ & Matrix elements & $\boldsymbol{\gamma_{ij}}$ & Matrix elements \\
        \midrule
        $\gamma_{11}$ & $ \frac{-g_1^2}{(4\pi)^2}$ & $\gamma_{17} = \gamma_{32} = \gamma_{87}$ & $\frac{6 y_u y_d}{(4\pi)^2}$  \\
        
        $\gamma_{22}$ & $\frac{-g_1^2}{(4\pi)^2} + \frac{9g_2^2}{(4\pi)^2} + \frac{8g_3^2}{(4\pi)^2} + \frac{3( (y_u)^2 + 2(y_d)^2)}{2(4\pi)^2}$ & $\gamma_{23} = \gamma_{58} = \gamma_{78} $ & $\frac{12 y_u y_d}{(4\pi)^2}$  \\
        
        $\gamma_{33}$ & $\frac{g_1^2}{2(4\pi)^2} + \frac{9g_2^2}{2(4\pi)^2} + \frac{8g_3^2}{(4\pi)^2} + \frac{3( (y_u)^2 + 2(y_d)^2)}{2(4\pi)^2}$ & $\gamma_{48} = \gamma_{63}= \gamma_{93}$ & $\frac{18 y_u y_d}{(4\pi)^2}$  \\
        
        $\gamma_{44}$ & $\frac{-4g_1^2}{(4\pi)^2} + \frac{3( (y_u)^2 + 2(y_d)^2)}{2(4\pi)^2}$ & $\gamma_{39} = \gamma_{84}$ & $\frac{3 y_u y_d}{(4\pi)^2}$  \\

        $\gamma_{55}$ & $\frac{2g_1^2}{(4\pi)^2}+ \frac{3( (y_u)^2 + 2(y_d)^2)}{2(4\pi)^2}$ & &  \\

        $\gamma_{66}$ & $\frac{2g_1^2}{(4\pi)^2} + \frac{8g_3^2}{(4\pi)^2} + \frac{3( (y_u)^2 + 2(y_d)^2)}{2(4\pi)^2}$ & &  \\
        
        $\gamma_{77}$ & $\frac{-g_1^2}{(4\pi)^2} +\frac{9 g_2^2}{(4\pi)^2}  + \frac{3( (y_u)^2 + 2(y_d)^2)}{2(4\pi)^2}$ & &  \\

        $\gamma_{88}$ & $\frac{g_1^2}{2(4\pi)^2} +\frac{9g_2^2}{2(4\pi)^2}  + \frac{3( (y_u)^2 + 2(y_d)^2)}{2(4\pi)^2}$ & &  \\

        $\gamma_{99}$ & $\frac{-4g_1^2}{(4\pi)^2} + \frac{8g_3^2}{(4\pi)^2} + \frac{3( (y_u)^2 + 2(y_d)^2)}{2(4\pi)^2}$ & &  \\
        \bottomrule
    \end{tabular}
    \caption{Matrix elements $\gamma_{ij}$ of the anomalous dimension matrix for energy scales below the new physics scale.}
    \label{table:AnomalousMatrix}
\end{table}

\subsection{Calculation of anomalous dimension in the presence of intermediate new physics field interactions}

In this section, we extend the anomalous dimension analysis to include possible interactions from intermediate-scale New Physics. We introduce:
\begin{itemize}
    \item $N$ scalar doublet fields,\footnote{In the Python code to compute the running~\cite{RepoName}, $N$ is defined as an integer input parameter that specifies the number of additional scalar doublets. The corresponding Yukawa couplings $y_{un}$ and $y_{dn}$ are initialized as arrays of length $N$.}
    \item $J$ $U(1)_Y$-like gauge interactions,\footnote{The code defines $J$ as the number of additional $U(1)$-like gauge groups. The corresponding gauge couplings $g_{1j}$ are stored in an array indexed by $j = 1, \dots, J$.}
    \item $K$ $SU(2)_L$-like gauge interactions,\footnote{$K$ is specified similarly in the code for $SU(2)_L$-like interactions. The gauge couplings $g_{2k}$ are implemented as an array indexed by $k = 1, \dots, K$.}
    \item $L$ $SU(3)_C$-like gauge interactions.\footnote{The code uses $L$ to denote the number of additional $SU(3)_C$-like interactions, with couplings $g_{3l}$ implemented as an array indexed by $l = 1, \dots, L$.}
\end{itemize}
These additional fields modify the RGE of the effective six-quark operators by altering the beta functions in the Callan-Symanzik equation.

The modified beta functions for gauge and Yukawa couplings take the following form:
\begin{align}
    \beta_{im} = - \frac{g_{im}}{16 \pi^2} \beta_{im}^0 \quad (i = 1, 2, 3: m = j,k,l), \quad \beta_{y_{i'{n}}} = - \frac{y_{i'{n}}}{16 \pi^2} \beta_{y_{i'{n}}}^0 \quad (i' = u,d) \label{eq:bfn_NP_Full}
\end{align}

where,
\begin{align}
    \beta_{1j}^0 &= - \frac{2}{3} (N_f) g^2_{1j} - \frac{g^2_{1j}(N+1)}{10}, \notag \\ 
    \beta_{2k}^ 0 &= (g^2_{2k})\Bigg[\frac{16}{3}- \frac{2}{3} (N_f ) - \frac{N+1}{6}\Bigg]  + 2\Bigg(\sum^K_{k=1}(g^2_{2k})+(g^2_{20})\Bigg), \notag \\  
    \beta_{3l}^0 &= (g^2_{3l})\Bigg[8 - \frac{2}{3} N_f\Bigg] + 3\Bigg(\sum^L_{l=1}(g^2_{3l})+(g^2_{30})\Bigg), \notag \\
    \beta _{y_{u0}} = \beta _{y_{u1}}=,,, = \beta _{y_{uN}} &=  \frac{3}{2} (Y_{u}^{\dagger} Y_{u} - Y_{d}^{\dagger} Y_{d}) + T + \sum^{N}_{n=1}\Bigg[  (\frac{9}{2}y^2_{un} + \frac{3}{2} y^2_{dn})\Bigg] - \Bigg(\frac{17}{20} \Bigg[\sum^J_{j=1}(g^2_{1j})+(g^2_{10})\Bigg] \notag \\ 
    &+ \frac{9}{4} \Bigg[\sum^K_{k=1}(g^2_{2k})+(g^2_{20})\Bigg] + 8 \Bigg[\sum^L_{l=1}(g^2_{3l})+(g^2_{30})\Bigg] \Bigg), \notag \\
    \beta _{y_{d0}} = \beta _{y_{d1}}=,,, = \beta _{y_{dN}} &= \frac{3}{2} (Y_{d}^{\dagger} Y_{d} - Y_{u}^{\dagger} Y_{u}) + T + \sum^{N}_{n=1}\Bigg[  (\frac{9}{2}y^2_{dn} + \frac{3}{2} y^2_{un})\Bigg] - \Bigg(\frac{1}{4} \Bigg[\sum^J_{j=1}(g^2_{1j})+(g^2_{10})\Bigg] \notag \\ 
    & + \frac{9}{4} \Bigg[\sum^K_{k=1}(g^2_{2k})+(g^2_{20})\Bigg] + 8 \Bigg[\sum^L_{l=1}(g^2_{3l})+(g^2_{30})\Bigg] \Bigg). \label{eq:bfn_NP}
\end{align}

Here, \( T = \text{Tr} \left( 3 Y_d^\dagger Y_d + 3 Y_u^\dagger Y_u \right) \) accounts for the trace over Yukawa couplings of all fermion generations.

With these modifications, we compute the anomalous dimension matrix for the full set of nine independent six-quark operators, including their mixing above the New Physics scale. The resulting diagonal and off-diagonal elements of the matrix are summarized in Tab.~\ref{table:anomalous_dim_NP}\footnote{The Python code to compute the running~\cite{RepoName}, numerically integrates the RG equations for all couplings and Wilson coefficients using the scipy library with the RK45 method. The evolution is piecewise, segmented by intermediate energy scales \(E_i\) where new fields are introduced. At each \(E_i\), the values of \(N\), \(J\), \(K\), and \(L\) are updated and the coupling arrays are extended accordingly. See the accompanying code documentation for implementation details as given in Sec.~\ref{sec:PythonCode} and in GitHub repo.~\cite{RepoName}.}.

\begin{table}[h]
    \centering
    \begin{tabular}{c|c}
        \toprule
        \multicolumn{2}{c}{\textbf{Diagonal Elements}} \\
\hline
\hline
        $\boldsymbol{\gamma_{ij}}$ & Matrix elements \\
\hline
\hline
        $\gamma_{11}$ & $ \sum^{J}_{j=0}\frac{-g_{1j}^2}{(4\pi)^2}$  \\
        
        $\gamma_{22}$ & $\sum^{J}_{j=0}\frac{-g_{1j}^2}{(4\pi)^2} + \sum^{K}_{k=0}\frac{9g_{2k}^2}{(4\pi)^2} + \sum^{L}_{l=0}\frac{8g_{3l}^2}{(4\pi)^2} + \sum^{N}_{n=0}\frac{3( (y_{un})^2 + 2(y_{dn})^2)}{2(4\pi)^2}$   \\
        
        $\gamma_{33}$ & $\sum^{J}_{j=0}\frac{g_{1j}^2}{2(4\pi)^2}  + \sum^{K}_{k=0}\frac{9g_{2k}^2}{2(4\pi)^2} + \sum^{L}_{l=0}\frac{8g_{3l}^2}{(4\pi)^2} + \sum^{N}_{n=0}\frac{3( (y_{un})^2 + 2(y_{dn})^2)}{2(4\pi)^2}$  \\
        
        $\gamma_{44}$ & $\sum^{J}_{j=0}\frac{-4g_{1j}^2}{(4\pi)^2}  + \sum^{N}_{n=0}\frac{3( (y_{un})^2 + 2(y_{dn})^2)}{2(4\pi)^2}$  \\

        $\gamma_{55}$ & $\sum^{J}_{j=0}\frac{2g_{1j}^2}{(4\pi)^2} + \sum^{N}_{n=0}\frac{3( (y_{un})^2 + 2(y_{dn})^2)}{2(4\pi)^2}$ \\

        $\gamma_{66}$ & $\sum^{J}_{j=0}\frac{2g_{1j}^2}{(4\pi)^2} + \sum^{L}_{l=0}\frac{8g_{3l}^2}{(4\pi)^2} + \sum^{N}_{n=0}\frac{3( (y_{un})^2 + 2(y_{dn})^2)}{2(4\pi)^2}$  \\
        
        $\gamma_{77}$ & $\sum^{J}_{j=0}\frac{-g_{1j}^2}{(4\pi)^2} + \sum^{K}_{k=0}\frac{9 g_{2k}^2}{(4\pi)^2}  + \sum^{N}_{n=0}\frac{3( (y_{un})^2 + 2(y_{dn})^2)}{2(4\pi)^2}$ \\

        $\gamma_{88}$ & $\sum^{J}_{j=0}\frac{g_{1j}^2}{2(4\pi)^2}  + \sum^{K}_{k=0}\frac{9g_{2k}^2}{2(4\pi)^2} + \sum^{N}_{n=0}\frac{3( (y_{un})^2 + 2(y_{dn})^2)}{2(4\pi)^2}$  \\

        $\gamma_{99}$ & $\sum^{J}_{j=0}\frac{-4g_{1j}^2}{(4\pi)^2} + \sum^{L}_{l=0}\frac{8g_{3l}^2}{(4\pi)^2}  + \sum^{N}_{n=0}\frac{3( (y_{un})^2 + 2(y_{dn})^2)}{2(4\pi)^2}$   \\[1ex]
\hline
        \multicolumn{2}{c}{\textbf{Off-Diagonal Elements}} \\[1ex]
\hline
\hline
        $\boldsymbol{\gamma_{ij}}$ & Matrix elements \\[1ex]
\hline
\hline

        $\gamma_{17} = \gamma_{32} = \gamma_{87}$ & $\sum^{N}_{n=0}\frac{6 y_{un} y_{dn}}{(4\pi)^2}$\\

        $\gamma_{23} = \gamma_{58} = \gamma_{78} $ & $\sum^{N}_{n=0}\frac{12 y_{un} y_{dn}}{(4\pi)^2}$\\

        $\gamma_{48} = \gamma_{63}= \gamma_{93}$ & $\sum^{N}_{n=0}\frac{18 y_{un} y_{dn}}{(4\pi)^2}$\\
        
        $\gamma_{39} = \gamma_{84}$ & $\sum^{N}_{n=0}\frac{3 y_{un} y_{dn}}{(4\pi)^2}$ \\
        
        \bottomrule
    \end{tabular}
    \caption{One-loop anomalous dimension matrix elements for six-quark operators above the intermediate-scale New Physics threshold. Diagonal and off-diagonal entries depend on the number and coupling strengths of new fields.}
    \label{table:anomalous_dim_NP}
\end{table}

\section{Estimating the New Physics Scale \(\Lambda\) in Chiral Perturbation Theory}
\label{sec:scaleChPT}
Higher-dimensional operators frequently capture the effects of heavy new physics through an expansion of the effective field theory (EFT) in the study of physics beyond the SM. These operators can mediate uncommon or prohibited processes at low energies, but they are inhibited by high-energy scale powers \(\Lambda\). The dimension-nine six-quark operator provided in the following Eq.~\ref{eq:op_1_LEFT}-\ref{eq:op_9_LEFT} is an example of such an operator. Each operator can be written with the coupling constant as: 

\begin{equation}
\mathcal{O}^{(7)} = \frac{\widetilde{C}^{(7)}}{\Lambda^5} \left(\bar{d}^c_{iL} u_{jL}\right)\left(\bar{d}^c_{kL} u_{lL}\right)\left(\bar{d}^c_{mR} d_{nR}\right) \Gamma^{(1)}_{ijklmn},
\end{equation}

where the dimensionless Wilson coefficient is denoted by \(\widetilde{C}^{(7)}\). Evaluating this operator's matrix elements between hadronic states, such as nucleons, is necessary to comprehend its physical implications. At low energies, however, a straightforward evaluation in terms of quark fields is inappropriate because quarks are contained inside hadrons. We use chiral perturbation theory (ChPT), which is the low-energy effective theory of quantum chromodynamics (QCD), to relate such operators to quantifiable observables.

The main concept is that the coefficients of these hadronic operators encode the impact of novel physics at energies well below \(\Lambda\). Bounds on the scale \(\Lambda\) can then be established by comparing experimental data with the expected size of an observable (using ChPT and hadronic matrix elements). Only two distinct nuclear matrix elements for neutron decay parameterized in terms of two real numbers $\alpha$ and $\beta$ exist because of parity conservation of strong interactions and isospin symmetry. These matrix elements are given by,
\begin{align}
    \langle 0|\epsilon^{abc} (\bar{d}^{\dagger}_a \bar{u}^{\dagger}_b) d_c | n^{(s)} \rangle &= \alpha P_L u^{(s)}_n , &
    \langle 0|\epsilon^{abc} (d_a u_b) d_c | n^{(s)} \rangle &= \beta P_L u^{(s)}_n ,
\end{align}

where $s$ represents the proton spin, $u^{(s)}_p$ represents the 4-component neutron spinor, $a,b,c$ represents the color indices, and $P_{L,R}$ represents the chirality projectors. Imposing isospin symmetry yields the matrix components with an initial state neutron with a 4-component spinor $u^{(s)}_n$. The ones for right-handed spinors can be related to parity conservation.
\begin{align}
    \langle 0|\epsilon^{abc} (d_a u_b) \bar{d}^{\dagger}_c | n^{(s)} \rangle &= -\alpha P_R u^{(s)}_n , &
    \langle 0|\epsilon^{abc} (\bar{d}^{\dagger}_a  \bar{u}^{\dagger}_b ) \bar{d}^{\dagger}_c   | n^{(s)} \rangle &= -\beta P_R u^{(s)}_n,
\end{align}
where the values of the $\alpha$ and $\beta$ constants are \cite{Yoo:2021gql, Aoki:1999tw}
\begin{equation}
    \beta = 0.01269(107)\ \text{GeV}^3 , \qquad \alpha = -0.01257(111)\ \text{GeV}^3 .
\end{equation}

In this context, the study of neutron-antineutron oscillations is a well-known application of ChPT. Baryon number violating interactions mediate these hypothetical transitions between a neutron and its antiparticle. Such transitions may be caused by the dimension-nine operator \(\mathcal{O}^{(7)}\) mentioned above.

The appropriate matrix element for operators such as \(\mathcal{O}^{(n)}\) is expressed as follows
\begin{equation}
\langle \bar{n} | \mathcal{O}^{(n)} | n \rangle = \mathcal{C}^{(n)}_{\text{had}} = \frac{|A_n| \widetilde{C}^{(n)}}{(\Lambda^{(n)})^5},
\end{equation}
where, $A_n$ are given in Tab.~\ref{table:tab1}.
\begin{table}[h!]
    \centering
    \begin{tabular}{c|c}
        \hline
        \hline
        \textbf{Operators} & $A_n$ \\
        \hline
        $\mathcal{O}^{(1)}$ & $\alpha^2$  \\
        \hline
        $\mathcal{O}^{(2)}$ & $\alpha\beta$  \\
        \hline
        $\mathcal{O}^{(3)}$ & $\alpha\beta$  \\
        \hline
        $\mathcal{O}^{(4)}$ & $\beta^2$  \\
        \hline
        $\mathcal{O}^{(5)}$ & $\beta^2$  \\
        \hline
        $\mathcal{O}^{(6)}$ & $\beta^2$  \\
        \hline
        $\mathcal{O}^{(7)}$ & $\alpha^2$  \\
        \hline
        $\mathcal{O}^{(8)}$ & $\alpha\beta$  \\
        \hline
        $\mathcal{O}^{(9)}$ & $\beta^2$ \\
        \hline
        \hline
    \end{tabular}
    \caption{$A_n$ values}
    \label{table:tab1}
\end{table}

The hadronic matrix element \(\mathcal{C}^{(n)}_{\text{had}}\) needs to be estimated using symmetry arguments in ChPT or assessed using non-perturbative methods such as lattice QCD. The transition amplitude and the oscillation time \(\tau_{n\bar{n}}\) are inversely correlated as,
\begin{equation}
\delta m = \frac{1}{\tau_{n\bar{n}}} \sim \frac{|A_n|  \widetilde{C}^{(n)}}{(\Lambda^{(n)})^5}.
\end{equation}

From this, we can solve for \(\Lambda\),
\begin{equation}
\Lambda^{(n)} \sim \left( |A_n| \frac{\widetilde{C}^{(n)}}{\delta m} \right)^{1/5}.
\end{equation}

SuperKamiokande has imposed a lower bound on the oscillation time, $\tau_{n\bar{n}} \gtrsim 10^8 $ seconds~\cite{Super-Kamiokande:2020bov}. If the hadronic matrix element has a typical value, such as \(\mathcal{C}_{\text{had}} \sim 10^{-4}~\text{GeV}^6\), we can estimate,

\begin{equation}
\delta m \lesssim 10^{-24}~\text{GeV} \quad \Rightarrow \quad \Lambda^{(n)} \gtrsim 500 \text{TeV}.
\end{equation}

Any novel physics that causes such a BNV must take place at energy scales higher than $500$ TeV.

\section{Renormalization group evolution }
\label{sec:scalelimit}
 RGE of baryon number violating operator is essential for relating low-energy theories to high-scale observations. A strong phenomenological description of neutron-antineutron oscillations, which are mediated by \(\Delta B = 2\) six-quark operators, requires the running of operator coefficients from the electroweak scale to a new physics scale. We calculate their one-loop anomalous dimensions within the SM to assess how well these operators perform. Our computation uses Fierz identities to preserve operator independence and dimensional regularization in the \(\overline{\text{MS}}\) scheme.

 We extend the study beyond the SM by taking into account new physics scenarios to assess the reliability of predictions for neutron-antineutron oscillation operators under RGE.  In particular, we study the impact of adding scalar doublets and $U(1)$ abelian gauge bosons at regular energy intervals. The beta functions and anomalous dimension matrices that control the evolution of the nine-dimensional $\Delta B = 2$ operators are modified by these extra fields.

\subsection{Renormalisation group evolution with only SM}

In this case, we analyze the RGE of the nine six-quark operators responsible for neutron-antineutron oscillations, using only SM interactions. These operators, labeled $\mathcal{O}^{(1)}$ through $\mathcal{O}^{(9)}$, differ in their chiral and color structures:

\begin{itemize}
    \item $\mathcal{O}^{(1)}-\mathcal{O}^{(3)}$: involve combinations of left-handed and/or right-handed quarks with antisymmetric color contractions.
    \item $\mathcal{O}^{(4)}-\mathcal{O}^{(6)}$: composed entirely of right-handed quark fields.
    \item $\mathcal{O}^{(7)}-\mathcal{O}^{(9)}$: arise from Higgs-induced interactions, written in a form consistent with SM gauge invariance.
\end{itemize}

Using Yukawa interactions and SM gauge bosons (gluons, $W^\pm$, $B^0$) and one-loop anomalous dimensions, we solve the coupled RG equations for the Wilson coefficients $\widetilde{C}^{(i)}(\mu)$. Evolution is carried out to the New physics scale with beginning circumstances given as $\widetilde{C}^{(i)}(90.19~\mathrm{GeV}) = 1$ for all $i$.

\begin{figure}[h]
    \centering
    \includegraphics[scale=0.50]{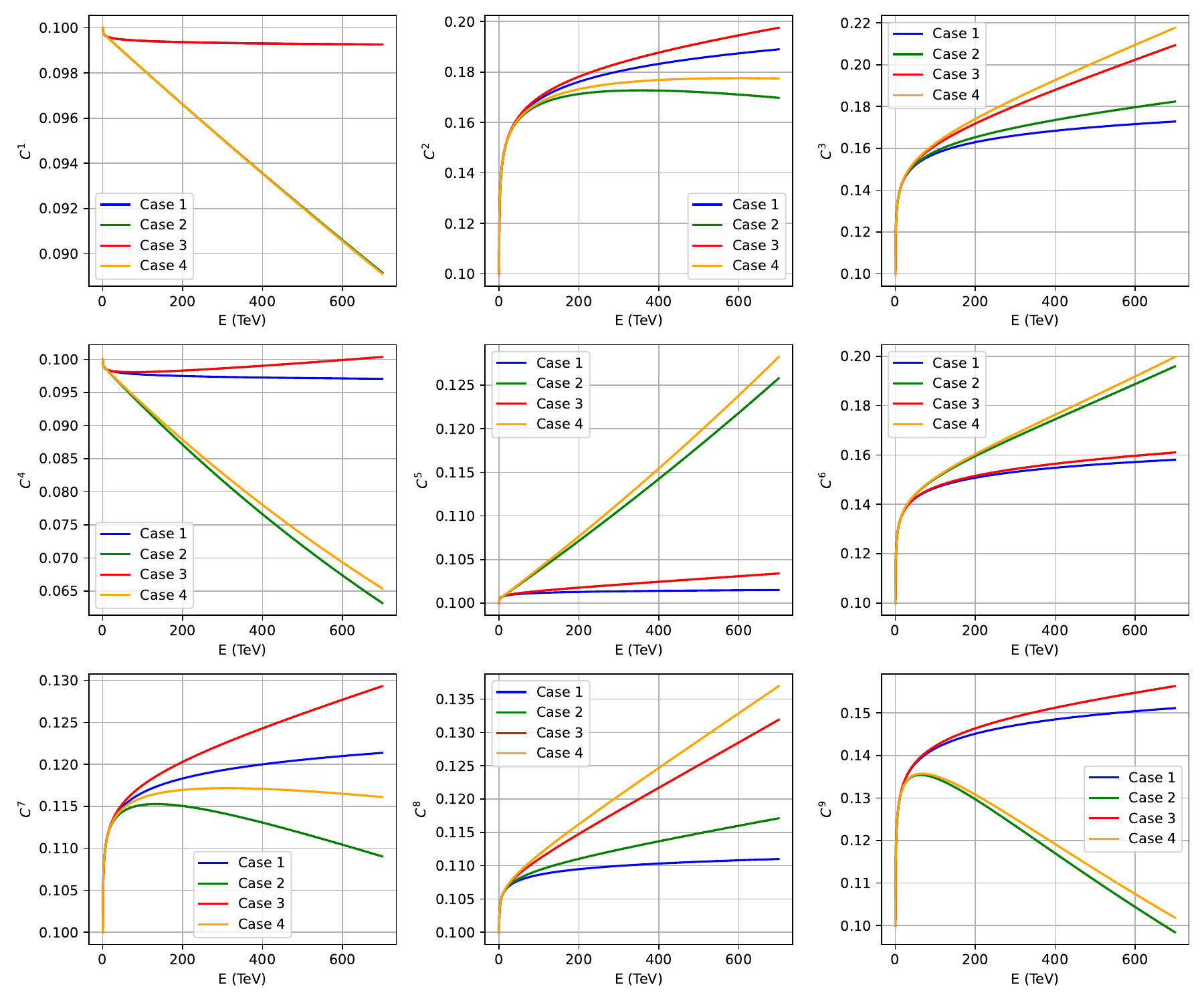}
    \caption{
Running of Wilson coefficients of all nine operators as functions of the energy scale $\mu$ under various initial conditions and parameter configurations:
Case 1(RGE with only SM) 
Case 2({RGE with $U(1)$ abelian gauge boson at every 10 TeV till 700 TeV})
Case 3({RGE with doublet scalar boson at every 10 TeV till 700 TeV})
Case 3({RGE with $U(1)$ abelian gauge boson and doublet scalar boson at every 10 TeV till 700 TeV}).
}
    \label{fig:rgewilson}
\end{figure}

The RGE of all nine Wilson coefficients, assuming only the SM, is plotted as a function of the energy scale $\mu$ as a blue curve in Fig.~\ref{fig:rgewilson}. 

\begin{figure}[h]
    \centering
    \begin{subfigure}[t]{0.5\textwidth}
        \centering
        \includegraphics[scale=0.52]{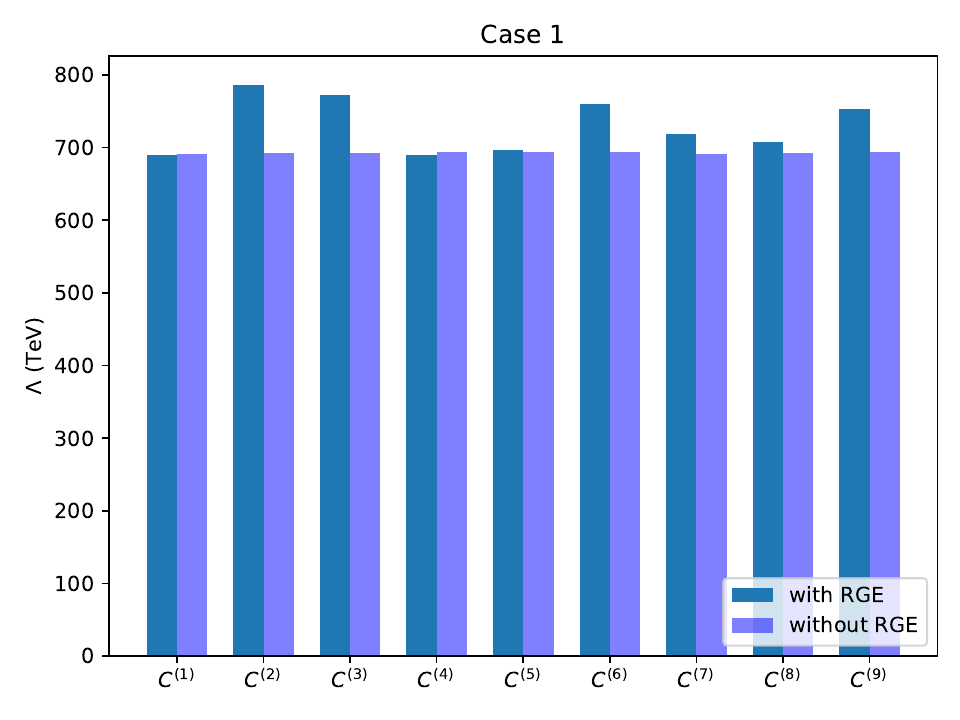}
        \caption{Limits on the UV scale}
        \label{fig:rgesm}
    \end{subfigure}%
    ~ 
    \begin{subfigure}[t]{0.5\textwidth}
        \centering
        \includegraphics[scale=0.4]{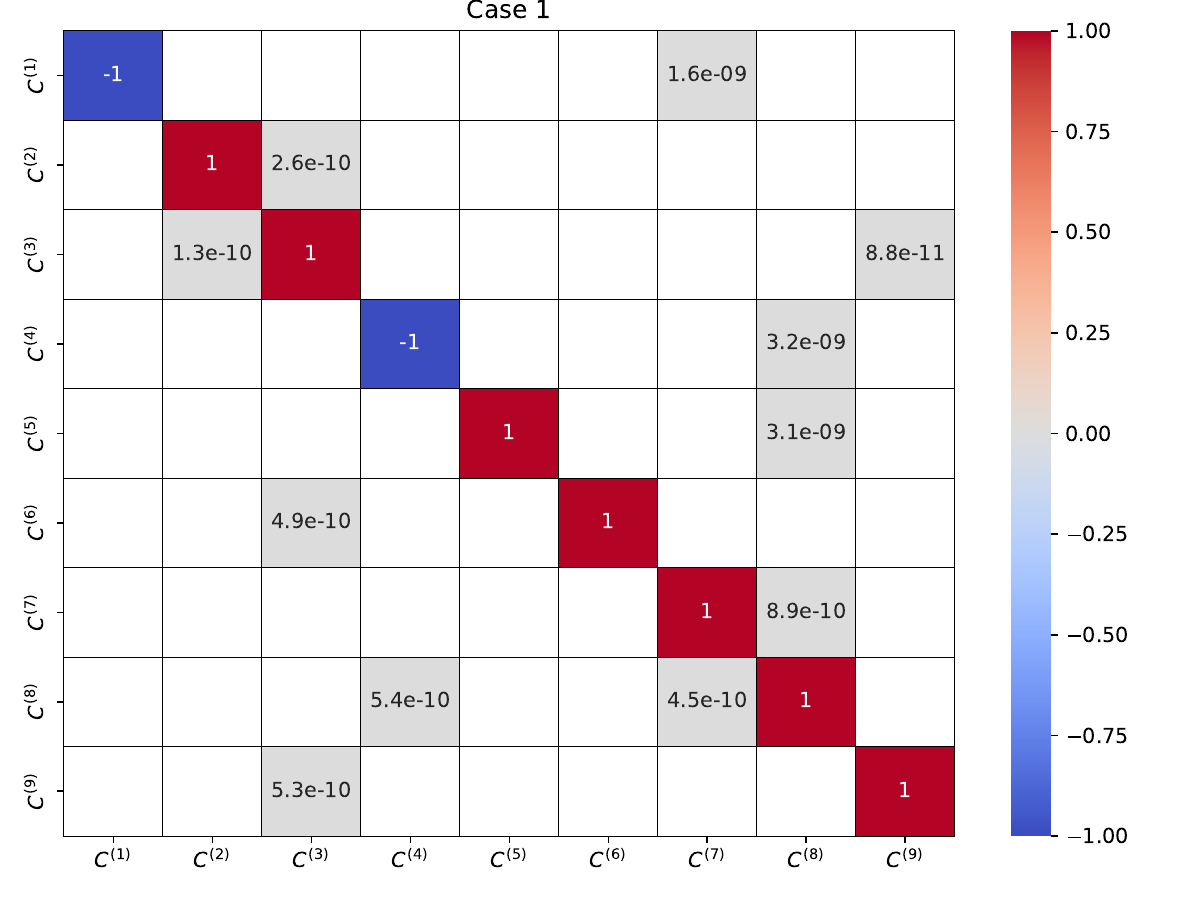}
        \caption{Correlation matrix at high scale}
            \label{fig:correlation_SM_700}
    \end{subfigure}
    \caption{(a) Limits on the UV scale $\Lambda^{(n)}$ for each of the nine operators without (with) RGE effects in dark blue(Violet) with only the SM. (b) Correlation matrix for dimension-9 six-quark operators under Standard Model RG evolution at the high scale $\mu = 700$ TeV.}
        \label{fig:correlationplots_SM}
\end{figure}

We calculate the updated bounds on the new physics scale $\Lambda^{(n)}$ for each of the nine six-quark operators by incorporating the effects of SM RGE. The Wilson coefficients $\widetilde{C}^{(i)}(\mu)$ are evolved from the electroweak scale ($\mu = 90.19~\mathrm{GeV}$) to a high-energy ultraviolet (UV) scale using the relevant RGE equations. These evolved coefficients yield modified constraints on $\Lambda^{(n)}$, which are compared to those obtained without RG running. The influence of renormalization group effects is illustrated in Fig.~\ref{fig:rgesm}.

The improvements are summarized in Tab.~\ref{tab:improved_limits_updated}. While $\Lambda^{(n)}_{\text{RGE}}$ incorporates the effect of one-loop running under SM interactions alone, $\Lambda^{(n)}_{\text{no RGE}}$ is computed without. The final column displays the relative improvement, emphasizing how crucial quantum corrections are when assessing limits on new physics.

\begin{table}[h!]
    \centering
    \begin{tabular}{c|c|c|c}
        \hline
        \hline
        \textbf{Operators} & $\Lambda^{(n)}_{\text{no RGE}}$ [TeV] & $\Lambda^{(n)}_{\text{RGE}}$ [TeV] & Improvement [\%] \\
        \hline
        $\mathcal{O}^{(1)}$ & 691 & 690 & -0.2 \\
        \hline
        $\mathcal{O}^{(2)}$ & 693 & 787 & 14 \\
        \hline
        $\mathcal{O}^{(3)}$ & 693 & 773 & 12 \\
        \hline
        $\mathcal{O}^{(4)}$ & 694 & 690 & -0.6 \\
        \hline
        $\mathcal{O}^{(5)}$ & 694 & 696 & 0.3 \\
        \hline
        $\mathcal{O}^{(6)}$ & 694 & 761 & 10 \\
        \hline
        $\mathcal{O}^{(7)}$ & 691 & 719 & 4 \\
        \hline
        $\mathcal{O}^{(8)}$ & 693 & 707 & 2 \\
        \hline
        $\mathcal{O}^{(9)}$ & 694 & 754 & 9 \\
        \hline
        \hline
    \end{tabular}
    \caption{
        Improvement in the limits on the new physics scale $\Lambda^{(n)}$ for each six-quark operator due to SM-only RGE effects (updated scenario). The third column shows the limits after RG running, and the last column shows the percentage improvement relative to the non-running case.
    }
    \label{tab:improved_limits_updated}
\end{table}


The correlation matrix plots, as shown in Fig.~\ref{fig:correlationplots_SM}, correspond to the SM scenario. They are computed from a normalised correlation matrix defined by
\begin{equation}
\rho_{ij}=\frac{\gamma_{ij}}{\sqrt{|\gamma_{ii}||\gamma_{jj}|}} \ ,
\label{eq:correlation}
\end{equation}
where $\gamma_{ij}$ denotes the entries of the anomalous dimension matrix for the dimension-9 six-quark operators.

\begin{figure}[h]
    \centering
    \includegraphics[scale=0.4]{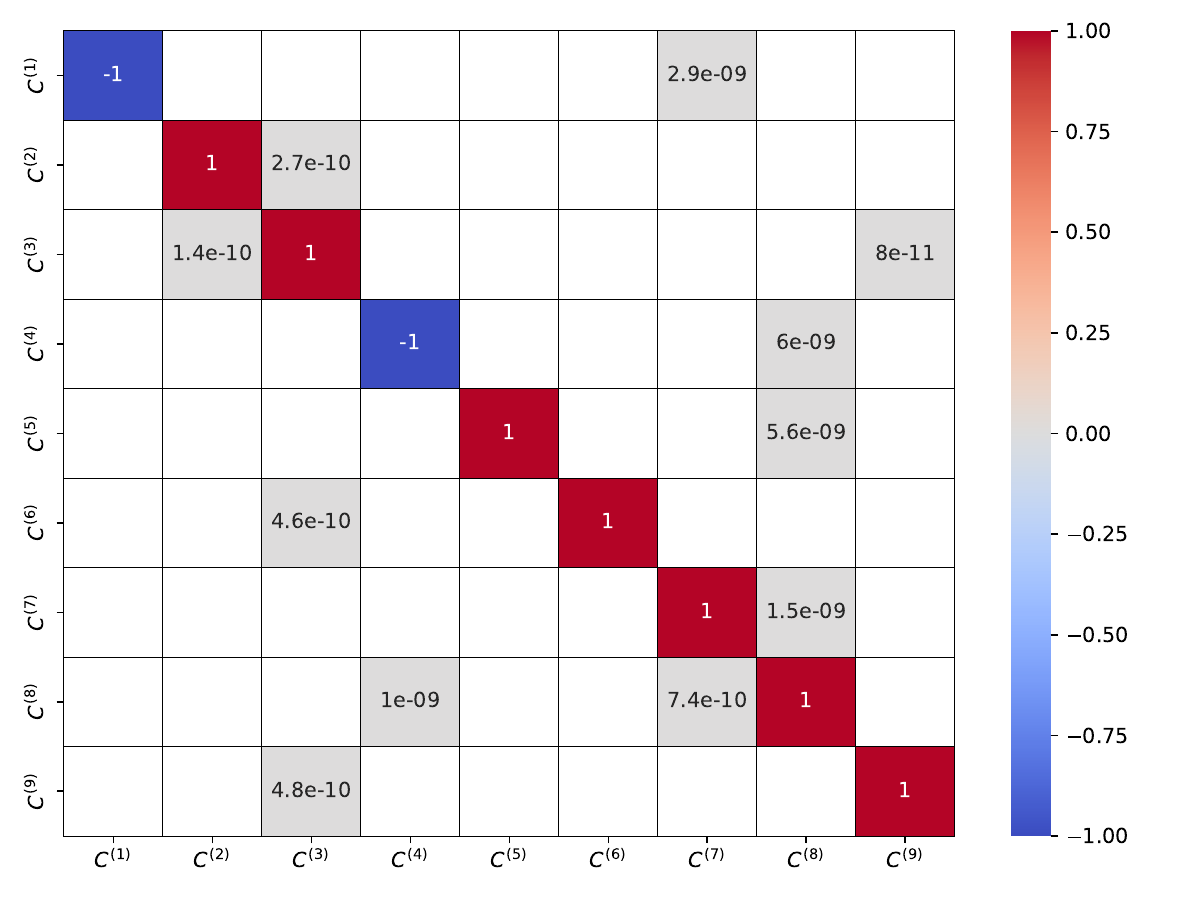}
    \caption{Correlation matrix $\rho$ for dimension-9 six-quark operators under SM at low scale.}
    \label{fig:correlation_SM_90}
\end{figure}

The matrix evaluated at $\mu = 90.19$~GeV (Fig.~\ref{fig:correlation_SM_90}) captures the initial structure at the electroweak scale, showing predominantly diagonal elements with minimal off-diagonal mixing. The high-scale matrix at $\mu = 700$~TeV (Fig.~\ref{fig:correlation_SM_700}) reflects the cumulative effects of SM running, where mild off-diagonal correlations appear due to operator mixing under purely SM dynamics.

\subsection{RGE with $U(1)$ abelian gauge boson new physics}

In this case, we take into account the insertion of a new $U(1)'_Y$ abelian gauge interaction, which corresponds to $J = 1, 2, \dots, 70$, at 10 TeV intervals up to 700 TeV. Through gauge corrections, each new gauge boson adds additively to the $\beta$-function and anomalous dimensions and relates similarly to hypercharge.

This leads to the modified running:
\begin{equation}
\beta_{1j} = \frac{g_{1j}^3}{16\pi^2} \left( \frac{2}{3} N_f + \frac{N + 1}{10} \right),
\end{equation}
And new diagonal entries in the anomalous dimension matrix acquire contributions of the form:
\begin{equation}
\gamma_{ii}^{(J)} \supset -\sum_{j=1}^{J} \frac{g_{1j}^2}{(4\pi)^2} \cdot \widetilde{C}^{(i)},
\end{equation}
where $\widetilde{C}^{(i)}$ are coefficients dependent on the hypercharge structure of each operator.

The green curve in Fig.~\ref{fig:rgewilson} shows the RG development of each of the nine Wilson coefficients under this expansion of the cumulative gauge. Several operators, particularly those sensitive to hypercharge mixing, are suppressed or enhanced more when more $U(1)$ fields are active at progressively larger scales.

\begin{figure}[h]
    \centering
    \begin{subfigure}[t]{0.5\textwidth}
        \centering
        \includegraphics[scale=0.52]{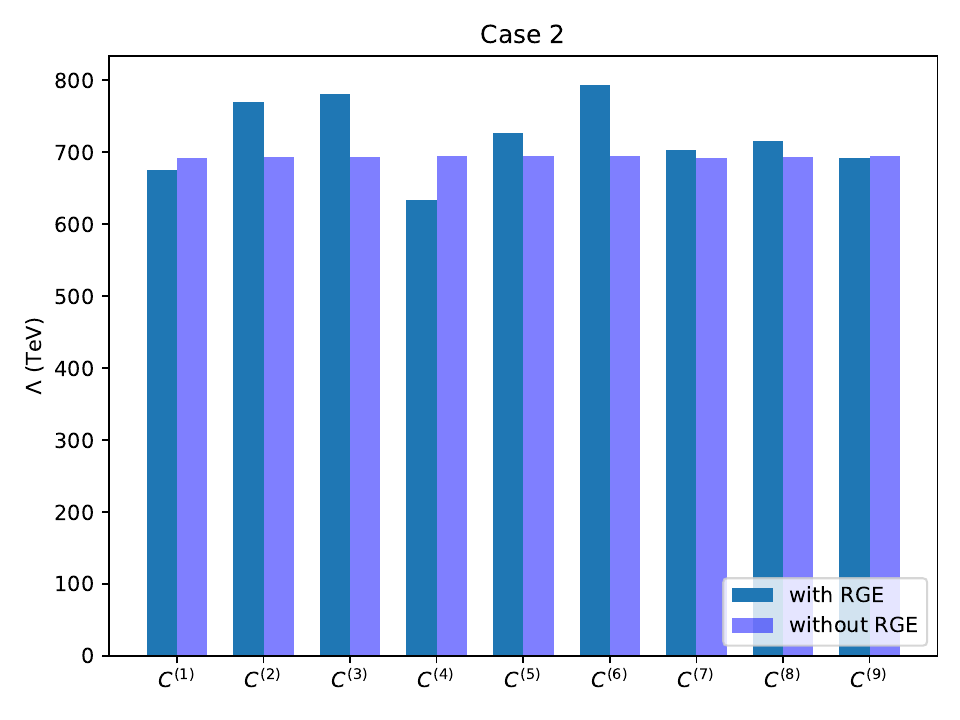}
        \caption{Limits on the UV scale}
        \label{fig:rgeU1}
    \end{subfigure}%
    ~ 
    \begin{subfigure}[t]{0.5\textwidth}
        \centering
        \includegraphics[scale=0.4]{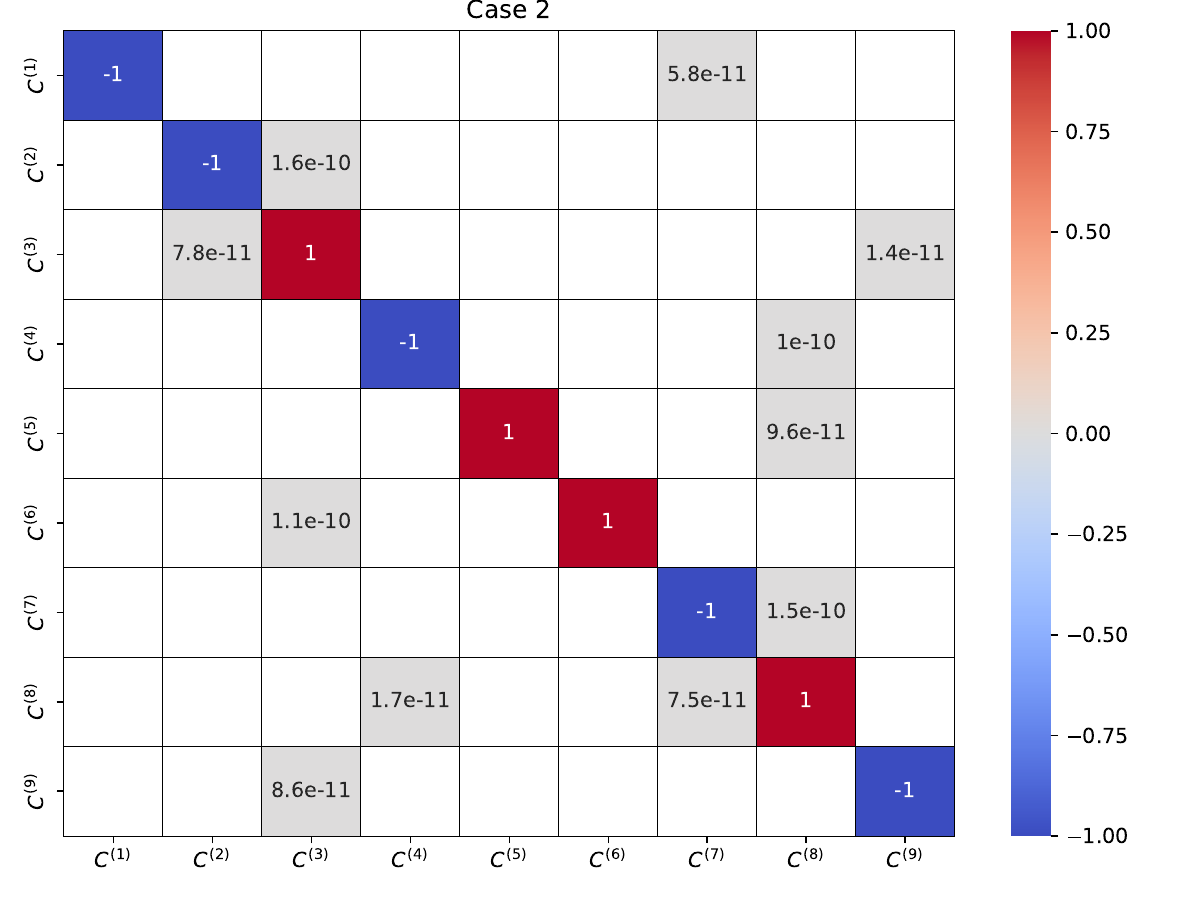}
        \caption{Correlation matrix }
            \label{fig:correlationplots_U1}
    \end{subfigure}
    \caption{(a) Limits on the UV scale $\Lambda^{(n)}$ for each of the nine operators without (with) RGE effects in dark blue(Violet) in the scenario with $U(1)$ gauge bosons at $10$~TeV onward. (b) Correlation matrix for dimension-9 six-quark operators in the presence of sequential $U(1)_Y$ gauge bosons at high scale $\mu = 700$ TeV.}
\end{figure}

We compute the ultraviolet (UV) scale limits $\Lambda^{(n)}$ for each of the nine operators with and without RG evolution, as shown in Fig.~\ref{fig:rgeU1}, in order to measure the effect of these successive $U(1)$ insertions on operator sensitivity. These limits, which are shown in Tab.~\ref{tab:improved_limits_U1}, clearly show that cumulative vector boson corrections have increased sensitivity. 

\begin{table}[h]
    \centering
    \begin{tabular}{c|c|c|c}
        \hline
        \hline
        \textbf{Operators} & $\Lambda^{(n)}_{\text{no RGE}}$ [TeV] & $\Lambda^{(n)}_{\text{RGE}}$ [TeV] & Improvement [\%] \\
        \hline
        $\mathcal{O}^{(1)}$ & 691 & 676 & -2  \\
        \hline
        $\mathcal{O}^{(2)}$ & 693 & 770 & 11 \\
        \hline
        $\mathcal{O}^{(3)}$ & 693 & 781 & 13 \\
        \hline
        $\mathcal{O}^{(4)}$ & 694 & 633 & -9 \\
        \hline
        $\mathcal{O}^{(5)}$ & 694 & 727 & 5 \\
        \hline
        $\mathcal{O}^{(6)}$ & 694 & 794 & 14  \\
        \hline
        $\mathcal{O}^{(7)}$ & 691 & 703 & 2 \\
        \hline
        $\mathcal{O}^{(8)}$ & 693 & 715 & 3 \\
        \hline
        $\mathcal{O}^{(9)}$ & 694 & 692 & -0.3 \\
        \hline
        \hline
    \end{tabular}
    \caption{
        Improvement in the UV scale limits $\Lambda^{(n)}$ for all nine operators under RGE evolution with sequential $U(1)$ abelian gauge bosons introduced every 10 TeV up to 700 TeV.
    }
    \label{tab:improved_limits_U1}
\end{table}

The correlation matrix plots, as shown in Fig.~\ref{fig:correlationplots_U1}, correspond to the scenario where only a new \( U(1)_Y \) gauge boson is introduced at each threshold from 10~TeV up to 700~TeV. We use the definition of the correlation matrix given in Eq.~\ref{eq:correlation}. This setup extends the SM by sequentially adding new abelian gauge bosons at discrete energy steps. At the electroweak scale $\mu = 90.19$~GeV (Fig.~\ref{fig:correlation_SM_90}), the matrix reflects the initial condition where diagonal terms dominate and operator mixing is minimal. By the high scale $\mu = 700$~TeV (Fig.~\ref{fig:correlationplots_U1}), the effects of gauge interactions have accumulated, where mild off-diagonal correlations appear due to operator mixing through the new $U(1)_Y$ sectors.

\subsection{RGE with doublet scalar boson new physics}

In this case, we create new electroweak scalar doublets, which are similar to the SM Higgs doublet, at intervals of 10 TeV, so that $N = 1, 2, \dots, 70$ by 700 TeV. Mostly through Yukawa-induced contributions, these new scalars influence the anomalous dimension matrix and interact with the SM quarks through Yukawa-like couplings.

The modified Yukawa $\beta$-functions now include:
\begin{equation}
\beta_{y_u} = \sum_{n=1}^{N} \left[ \frac{9}{2} y_{un}^2 + \frac{3}{2} y_{dn}^2 \right] - \text{gauge terms},
\end{equation}
with corresponding anomalous dimension matrix contributions:
\begin{align}
\gamma_{ii}^{(N)} &\supset \sum_{n=1}^{N} \frac{3\left(y_{un}^2 + 2 y_{dn}^2\right)}{2(4\pi)^2}, \\
\gamma_{ij}^{\text{off-diag}} &\supset \sum_{n=1}^{N} \frac{k_{ij}^{(n)} y_{un} y_{dn}}{(4\pi)^2}.
\end{align}

where $k_{ij}^{(n)}$ are constants calculated through the loop calculations.

Notably, adding scalar doublets improves operator mixing and causes non-trivial corrections, particularly for operators that involve electroweak symmetry-breaking structures or interact with both left- and right-handed quark fields. Consequently, even at intermediate scales, a number of Wilson coefficients exhibit fast and nonlinear evolution with energy, deviating greatly from their SM behaviour.

The red curve in Fig.~\ref{fig:rgewilson} depicts the evolution of these coefficients, revealing significant differences above a few hundred TeV. These patterns highlight how baryon number violating operators are sensitive to extended scalar sectors, which can transform their high-energy behaviour significantly and impact limitations on related new physics scales.

\begin{figure}[h]
    \centering
    \begin{subfigure}[t]{0.5\textwidth}
   \centering
    \includegraphics[scale=0.52]{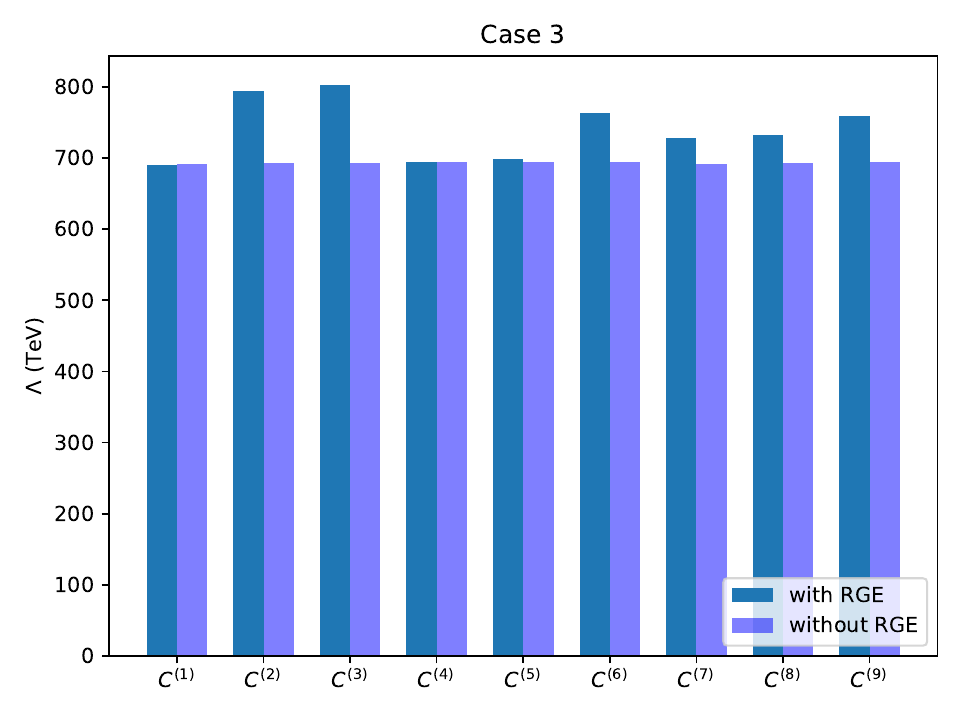}
    \caption{
Limits on the UV scale}
    \label{fig:rgeSD}
    \end{subfigure}%
    \begin{subfigure}[t]{0.5\textwidth}
     \centering
        \includegraphics[scale=0.4]{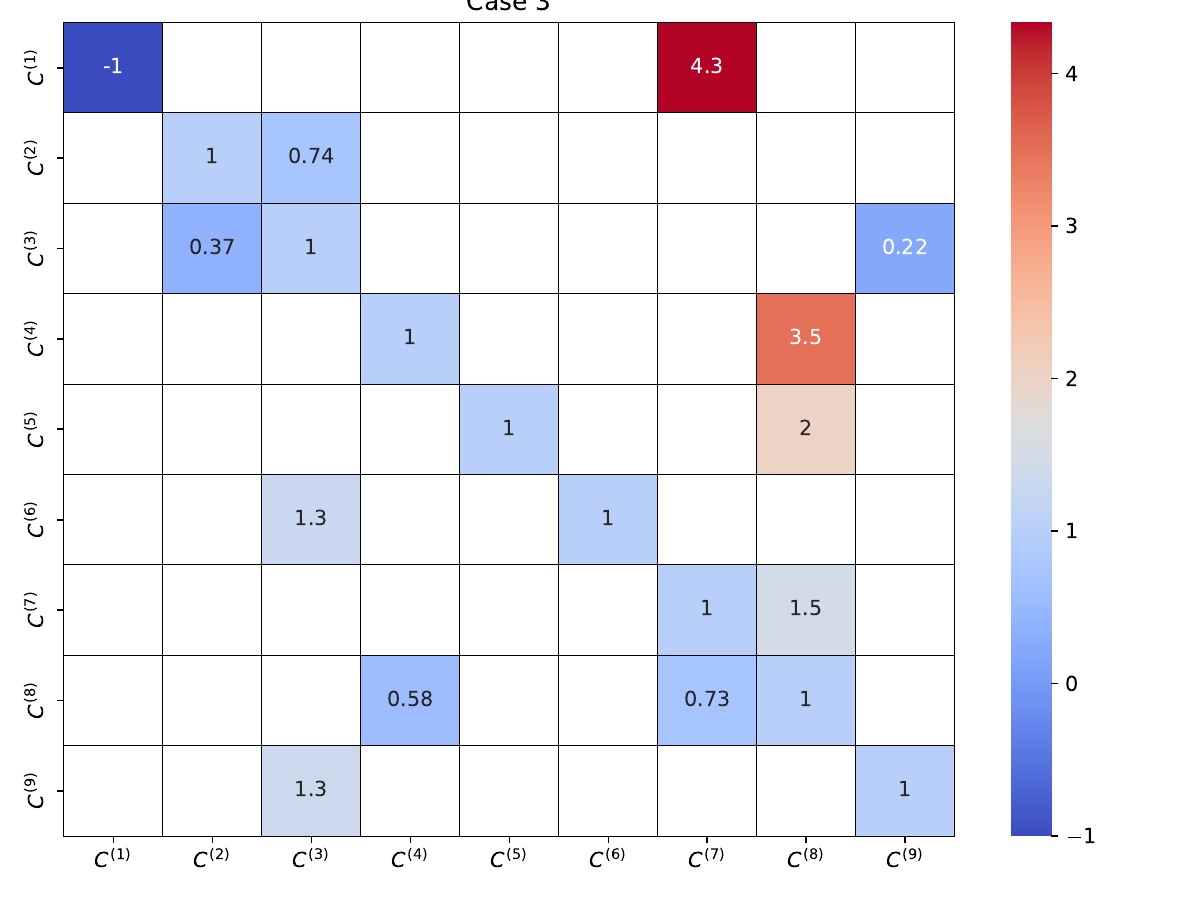}
        \caption{Correlation matrix $\rho$ at high scale.}
        \label{fig:correlationplots_SD}
    \end{subfigure}
    \caption{(a) Limits on the UV scale $\Lambda^{(n)}$ for each of the nine operators without (with) RGE effects in dark blue (Violet) in the scenario with scalar doublet at $10$~TeV onward. (b) Correlation matrix for dimension-9 six-quark operators in the presence of $U(1)$ gauge bosons and scalar doublets New Physics corresponds to the high scale $\mu = 700$ TeV.}
\end{figure}

The quantitative impact of these scalar doublet insertions on the new physics sensitivity for all nine operators is shown in Fig.~\ref{fig:rgeSD} and is summarized in Tab.~\ref{tab:improved_limits_SD}. For operators susceptible to electroweak symmetry breaking, the amplification is especially substantial and results from Yukawa-induced running and mixing.

\begin{table}[h]
    \centering
    \begin{tabular}{c|c|c|c}
        \hline
        \hline
        \textbf{Operators} & $\Lambda^{(n)}_{\text{no RGE}}$ [TeV] & $\Lambda^{(n)}_{\text{RGE}}$ [TeV] & Improvement [\%] \\
        \hline
        $\mathcal{O}^{(1)}$ & 691 & 690 & -0.2 \\
        \hline
        $\mathcal{O}^{(2)}$ & 693 & 794 & 15 \\
        \hline
        $\mathcal{O}^{(3)}$ & 693 & 803 & 16 \\
        \hline
        $\mathcal{O}^{(4)}$ & 694 & 695 & 0.14 \\
        \hline
        $\mathcal{O}^{(5)}$ & 694 & 699 & 0.7 \\
        \hline
        $\mathcal{O}^{(6)}$ & 694 & 763 & 10 \\
        \hline
        $\mathcal{O}^{(7)}$ & 691 & 728 & 5 \\
        \hline
        $\mathcal{O}^{(8)}$ & 693 & 732 & 6 \\
        \hline
        $\mathcal{O}^{(9)}$ & 694 & 759 & 9 \\
        \hline
        \hline
    \end{tabular}
    \caption{
       Improvements in $\Lambda^{(n)}$ limits for all nine operators with electroweak scalar doublets added at every 10 TeV up to 700 TeV.
    }
    \label{tab:improved_limits_SD}
\end{table}

The correlation matrix plots, as shown in Fig.~\ref{fig:correlationplots_SD}, correspond to the scenario where only a new scalar \( SU(2)_L \) doublet is introduced at each threshold from 10~TeV up to 700~TeV. The correlation matrix is defined in Eq.~\ref{eq:correlation}. In this setup, the SM is extended sequentially with new scalar doublets added at discrete energy thresholds. The matrix at $\mu = 90.19$~GeV (Fig.~\ref{fig:correlation_SM_90}) shows the initial operator structure at the electroweak scale, where diagonal elements dominate and off-diagonal correlations are negligible. At the high scale of $\mu = 700$~TeV (Fig.~\ref{fig:correlationplots_SD}), RG evolution with scalar-induced Yukawa mixing effects leads to enhanced off-diagonal correlations, reflecting the cumulative impact of repeated scalar insertions.

\subsection{RGE with $U(1)$ abelian gauge and doublet scalar boson new physics}

The two earlier expansions are combined in this complete scenario: a new $U(1)$ abelian gauge boson and a scalar doublet are introduced at each 10 TeV step from 10 TeV to 700 TeV, i.e., both $J = N = 1, 2, \dots, 70$.

This combined scenario produces the most significant departure from SM only RGE, as seen in the orange curve of Fig.~\ref{fig:rgewilson}. The combination of Yukawa driven enhancement and gauge induced suppression results in non-trivial behavior, with some coefficients growing greatly while others decrease quickly. Their gauge and chiral architectures are reflected in these operator-dependent patterns.


\begin{table}[!h]
    \centering
    \begin{tabular}{c|c|c|c}
        \hline
        \hline
        \textbf{Operators} & $\Lambda^{(n)}_{\text{no RGE}}$ [TeV] & $\Lambda^{(n)}_{\text{RGE}}$ [TeV] & Improvement [\%] \\
        \hline
        $\mathcal{O}^{(1)}$ & 691 & 676 & -2 \\
        \hline
        $\mathcal{O}^{(2)}$ & 693 & 777 & 12 \\
        \hline
        $\mathcal{O}^{(3)}$ & 693 & 809 & 17 \\
        \hline
        $\mathcal{O}^{(4)}$ & 694 & 638 & -8 \\
        \hline
        $\mathcal{O}^{(5)}$ & 694 & 729 & 5 \\
        \hline
        $\mathcal{O}^{(6)}$ & 694 & 797 & 15 \\
        \hline
        $\mathcal{O}^{(7)}$ & 691 & 712 & 3 \\
        \hline
        $\mathcal{O}^{(8)}$ & 693 & 738 & 6 \\
        \hline
        $\mathcal{O}^{(9)}$ & 694 & 697 & 0.4 \\
        \hline
        \hline
    \end{tabular}
    \caption{
        Improvement in the new physics scale limits $\Lambda^{(n)}$ for each six-quark operator when both $U(1)$ gauge bosons and scalar doublets are introduced at every 10 TeV step up to 700 TeV.
    }
    \label{tab:improved_limits_U1SD}
\end{table}

\begin{figure}[h]
    \centering
    \begin{subfigure}[t]{0.5\textwidth}
        \centering
        \includegraphics[scale=0.52]{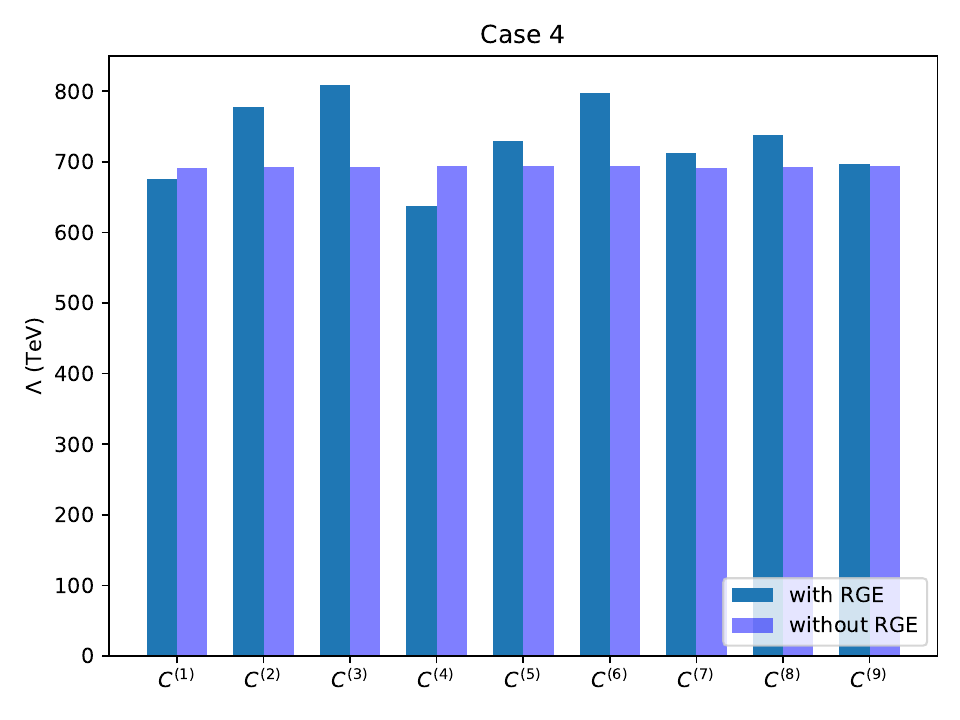}
        \caption{Limits on the UV scale}
        \label{fig:rgeU1SD}
    \end{subfigure}%
    ~ 
    \begin{subfigure}[t]{0.5\textwidth}
        \centering
        \includegraphics[scale=0.4]{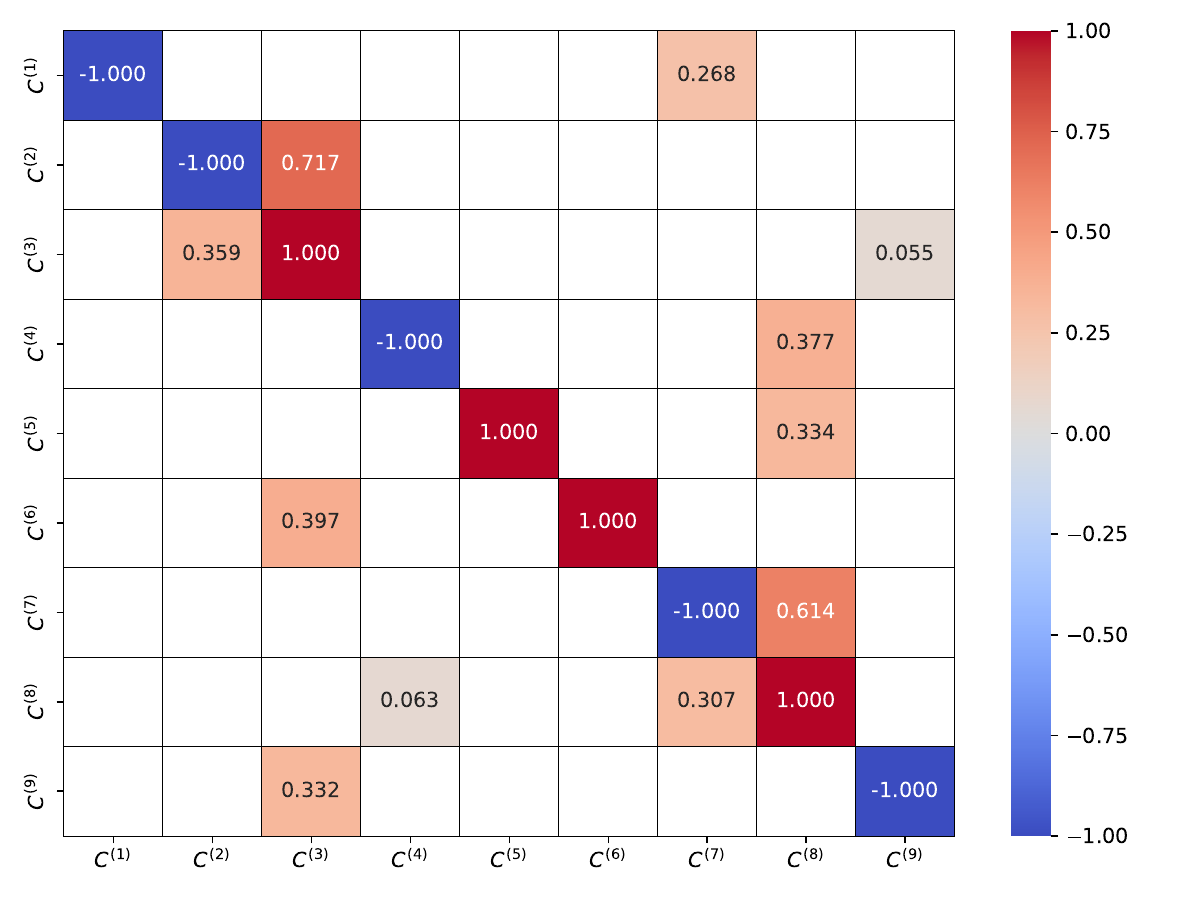}
        \caption{Correlation matrix }
            \label{fig:correlationplots_U1SD}
    \end{subfigure}
    \caption{(a) Limits on the UV scale $\Lambda^{(n)}$ for each of the nine operators without (with) RGE effects in dark blue(Violet) in the scenario with both $U(1)$ gauge bosons and scalar doublet are present at $10$~TeV onward. (b) Correlation matrix for dimension-9 six-quark operators in the presence of $U(1)$ gauge bosons and scalar doublets New Physics corresponds to the high scale $\mu = 700$ TeV.}
\end{figure}

For each of the nine operators, we compute and show the limits on $\Lambda^{(n)}$ in Fig.~\ref{fig:rgeU1SD} and in Tab.~\ref{tab:improved_limits_U1SD}. The combined effects of scalar-induced Yukawa enhancement and gauge suppression are reflected in the reported improvements.

The correlation matrix plot, as shown in Fig.~\ref{fig:correlationplots_U1SD}, correspond to the scenario where both a new \( U(1)_Y \) gauge boson and a scalar \( SU(2)_L \) doublet are introduced at each 10~TeV step from 10~TeV up to 700~TeV. 
The definition of the correlation matrix is given in Eq.~\ref{eq:correlation}. This setup involves a sequential extension of the SM with new fields at discrete thresholds. The matrix evaluated at 90.19~GeV (Fig.~\ref{fig:correlation_SM_90}) represents the initial structure of the operator coefficients at the electroweak scale, where only diagonal elements are prominent and off-diagonal contributions are negligible. In contrast, the matrix at 700\,TeV (Fig.~\ref{fig:correlationplots_U1SD}) reflects the result of RG evolution up to the high scale, capturing the cumulative effects of operator mixing and quantum corrections across all intermediate thresholds.

\section{Introduction to the Python code}
\label{sec:PythonCode}
The Python code numerically solves the differential equations for the Wilson coefficients as defined in Eq.~\ref{eq:WC_An_dim}, where the anomalous dimension matrix $\gamma$ is given in Tab.~\ref{table:anomalous_dim_NP}, and the $\beta$-functions are given in Eq.~\ref{eq:bfn_NP_Full} and Eq.~\ref{eq:bfn_NP}. The evolution of the Wilson coefficients is coupled to the running of the couplings $y_{un}$, $y_{dn}$, $g_{1j}$, $g_{2k}$, and $g_{3l}$, therefore requiring the simultaneous solution of the differential equations for all these quantities. The integration is performed using the RK45 solver from the scipy library.

The code requires the following inputs:
\begin{itemize}
\item The initial energy scale ($E_{\text{ini}}$) from which the evolution begins.
\item The initial values of the Wilson coefficients at $E_{\text{ini}}$ ($C_{\text{ini}}$).
\item The intermediate energy scales $E_i$ at which new couplings are introduced.
\item The final energy scale ($E_{\text{fin}}$) up to which the evolution is performed.
\item The values of $N$, $J$, $K$, and $L$ in each energy region, defining the structure of the couplings.
\item The values of the new couplings introduced at each $E_i$.
\end{itemize}

The code solves the coupled differential equations over each energy region, and at every $E_i$, it introduces the new couplings and updates the initial conditions for the next region. The integration is performed on a logarithmic scale, where $t = \ln(\mu)$.

The output consists of:

t: a list where each element corresponds to a specific energy region and contains the values of $t = \ln(\mu)$ within that region.

y: a list where each element is an array of the corresponding solutions for the couplings and Wilson coefficients at the values of t in the same region.

The solution `y' is returned as a dictionary with the following keys:

`g1', `'g2', `g3': arrays of gauge coupling values,

`yu', `yd': arrays of Yukawa coupling values,

`wc': array of Wilson coefficient values.

Each key maps to the corresponding evolution of the quantity over the specified energy range.
\section{Summary}
\label{sec:summary}
With the observation of 11 possible $n-\bar{n}$ oscillation candidates with an expected background of $9.3\pm2.7$ events at Super-Kamiokande and the prediction of new physics within the energy range of future collider searches, the BNV by two units has become interesting. Moreover, they provide a viable low-scale baryogenesis scenario, while avoiding stringent constraints from nucleon decay bounds. Due to this, dedicated experiments have been proposed to detect the neutron-antineutron and other dinucleon decay signatures. With these upcoming detectors, it is important to reliably predict the lifetime of these processes, accounting for the renormalization group evolution of the respective Wilson coefficients. In conjunction with the fact that {\it Large Hadron Collider} has not ruled out any new physics, with interaction with light quarks, above $\sim 10 ~\text{TeV}$ scale, does imply that these intermediate new physics scenarios can play a crucial role in the running of the BNV Wilson coefficients.

In this article, we compute the RG running of these coefficients under two assumptions. First, assuming that there exists only SM uptill the BNV scale, and the other, assuming that there exists bosonic new physics beyond the $10~\text{TeV}$ scale that influences the running. Though any number of such intermediate new physics scenarios can be envisaged, we study the modifications to the RG evolution in the presence of a doublet scalar and an abelian vector boson. The running is shown in Fig.~\ref{fig:rgewilson} and the new limits on the Wilson coefficients are shown in Fig.~\ref{fig:rgesm}, Fig.~\ref{fig:rgeU1}, Fig.~\ref{fig:rgeSD}, and Fig.~\ref{fig:rgeU1SD}. And we have summarized our results in Tab.~\ref{tab:improved_limits_updated}, Tab.~\ref{tab:improved_limits_U1}, Tab.~\ref{tab:improved_limits_SD}, and Tab.~\ref{tab:improved_limits_U1SD} respectively. Moreover, we have given the correlation matrix for each of these scenarios in Fig.~\ref{fig:correlation_SM_700}, Fig.~\ref{fig:correlationplots_U1}, Fig.~\ref{fig:correlationplots_SD} and Fig.~\ref{fig:correlationplots_U1SD}. The analysis can be generalized easily to any new physics beyond the electroweak scale and with $\Delta B=1$ operators~\cite{Beneito:2023xbk}. An improved study of BNV with Standard Model Effective Field Theory operators and their matching is in preparation. To that effort, we have used a Python script to RG evolve the Wilson coefficients, and it is attached along with this article~\cite{RepoName}. This code can be modified easily to study any bosonic new physics contribution.

\section*{Acknowledgments}
M.T.A. acknowledges the financial support of DST through the INSPIRE Faculty grant DST/INSPIRE/04/2019/002507. 
\section*{Appendix}

\appendix

\section{One-Loop Yukawa Corrections}
\label{sec:AppendixA}
In this appendix, we present the complete one-loop Yukawa correction calculation for diagrams (a3) and (d3) of the operator \(\mathcal{O}^{(2)}\), carried out within the \(\overline{\text{MS}}\) renormalization scheme.

\subsection{Fermion Self-Energy at One-Loop}

The full 1PI (one-particle irreducible) two-point function at one-loop includes the tree-level term, loop correction, and corresponding counterterms:
\begin{align}
    i \Gamma(\not{p}) = i(\not{p} - M) - i \Sigma(\not{p}) + i (\delta_{Z_\psi} \not{p} - \delta_M).
\end{align}
where \(\Sigma(\not{p})\) denotes the self-energy, and \(\delta_{Z_\psi}\), \(\delta_M\) are the wavefunction and mass counterterms, respectively.

The renormalization conditions at zero external momentum are given by,
\begin{align}
    i \Sigma(0) + \delta_M = 0 \\
    {\frac{d}{d\not{p}}\Sigma(\not{p})}\Bigr\rvert_{\not{p} = 0} - \delta_{Z_\varphi} = 0
\end{align}

\begin{figure}[h]
    \centering
    \includegraphics[scale=0.15]{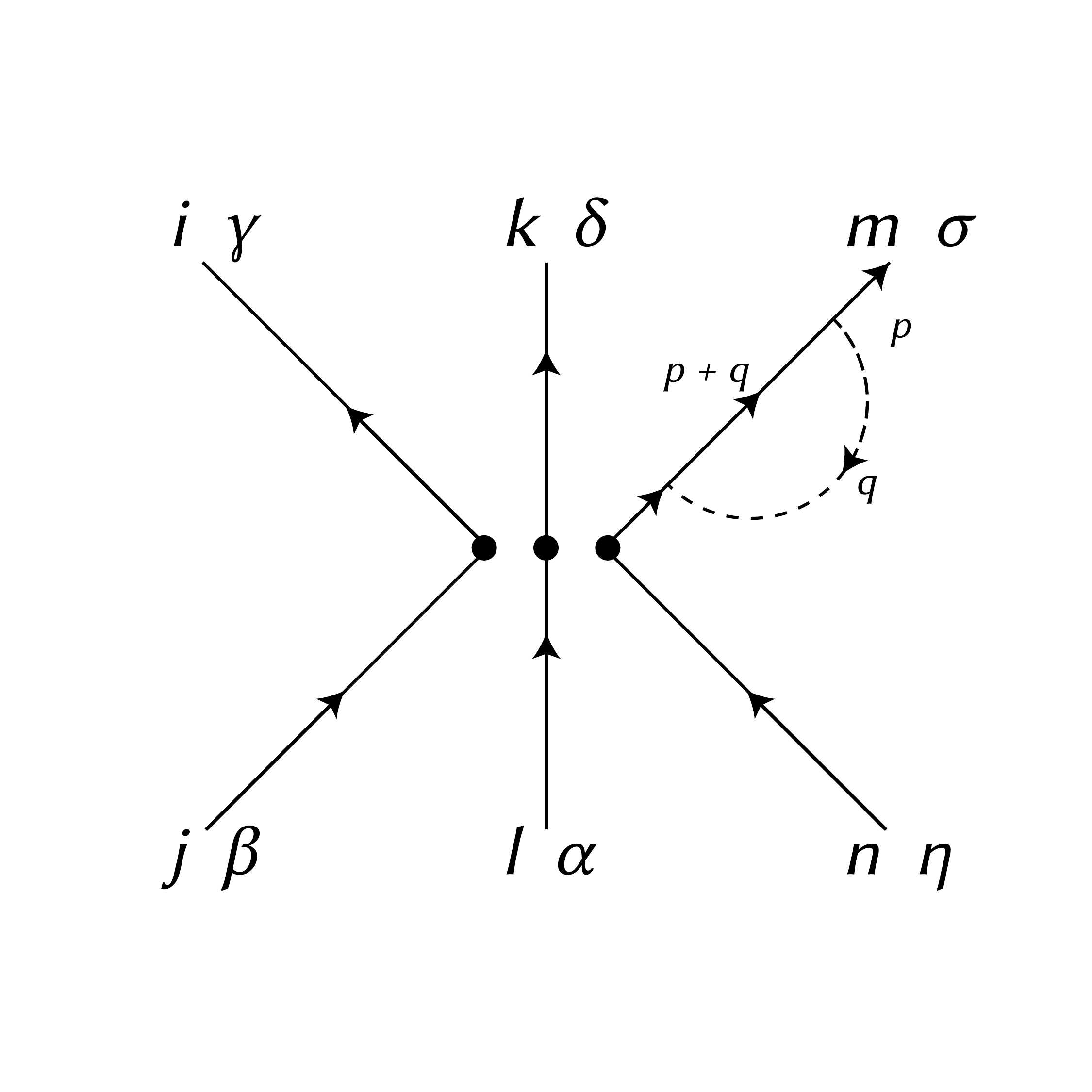}
    \caption{One-loop fermion self-energy diagram for Yukawa interaction.}
    \label{fig:Ap_A_Yukawa_self_One_loop}
\end{figure}

We compute the wavefunction renormalization constant \(\delta Z_\psi\) by evaluating the self-energy diagram using dimensional regularization, as shown in Fig.~\ref{fig:Ap_A_Yukawa_self_One_loop}. Here, \(m\) and \(M\) denote the masses of the scalar (Higgs) and the fermion, respectively.

The loop integral is
\begin{align}
    - i \Sigma(\not{p}) &= \mu^{\epsilon}g^2 \int \frac{d^d q}{(2\pi)^d} \frac{i}{q^2 - m^2}  \frac{i(\not{q} + \not{p} - M)}{(q + p)^2 - M^2}  \\
    &= -\mu^{\epsilon}g^2 \int \frac{d^d q}{(2\pi)^d} \frac{1}{q^2 - m^2} \frac{-\not{q} - \not{p} + M}{(q + p)^2 - M^2}
\end{align}


The standard dimensional regularization integral yields:
\begin{align}
    \int \frac{d^d q}{(2\pi)^d} \frac{1}{[q^2 - \Delta]^2} &= i \frac{\Gamma(\epsilon)}{(4\pi)^{\epsilon}} \frac{1}{\Delta^{\epsilon}},
\end{align}

Where, 

$\Delta = - x(1-x)p^2 + (1-x)m^2 + xM^2$.

Substituting back:
\begin{align}
    - i \Sigma(\not{p}) &= -\mu^{\epsilon}g^2 \int_0^1 dx \big[-(1-x) \not{q} + M\big] \frac{i \Gamma(\epsilon)}{(4\pi)^{2-\epsilon}} \frac{1}{\big[-x(1-x)p^2 + (1-x)m^2 + xM^2\big]^{\epsilon}}.
\end{align}

Expanding the terms in $\epsilon$:
\begin{align}
    - i \Sigma(\not{p}) &= -\mu^{\epsilon}g^2 \int_0^1 dx  \big[-(1-x) \not{q} + M\big] \frac{i}{(4\pi)^2} \ln\bigg[ \frac{-x(1-x)p^2 + (1-x)m^2 + xM^2}{\mu^2}\bigg].
\end{align}

The wavefunction counterterm is extracted via:
\begin{align}
\delta Z_\psi &= \left. \frac{d \Sigma(\not{p})}{d \not{p}} \right|_{\not{p} = 0} = -g^2 \frac{1}{(4\pi)^2} \frac{1}{\epsilon} \cdot \frac{1}{2} + \text{finite}.
\end{align}

\subsection{Yukawa Vertex Correction at One-Loop}
The renormalized 1PI vertex function at one-loop is:
\begin{align}
-i\Gamma(p_f, p_i) &= \widetilde{C}^{(2)} - iV(p_f, p_i) + \delta Z^{\phi}_{\mathcal{O}^{(2)}(d3)}
\end{align}

Where \(V(p_f, p_i)\) denotes the one-loop correction and \(\delta g\) is the vertex counterterm. At zero external momenta:
\begin{align}
-i\Gamma(0, 0) &= \widetilde{C}^{(2)} - iV(0, 0) + \delta Z^{\phi}_{\mathcal{O}^{(2)}(d3)} = \widetilde{C}^{(2)}\\
\delta Z^{\phi}_{\mathcal{O}^{(2)}(d3)} &= -iV(0, 0)
\end{align}

We now compute the counterterm for the Fig.~\ref{fig:Ap_A_Yukawa_vertex_One_loop}, using dimensional regularization:

\begin{figure}[h]
    \centering
    \includegraphics[scale=0.15]{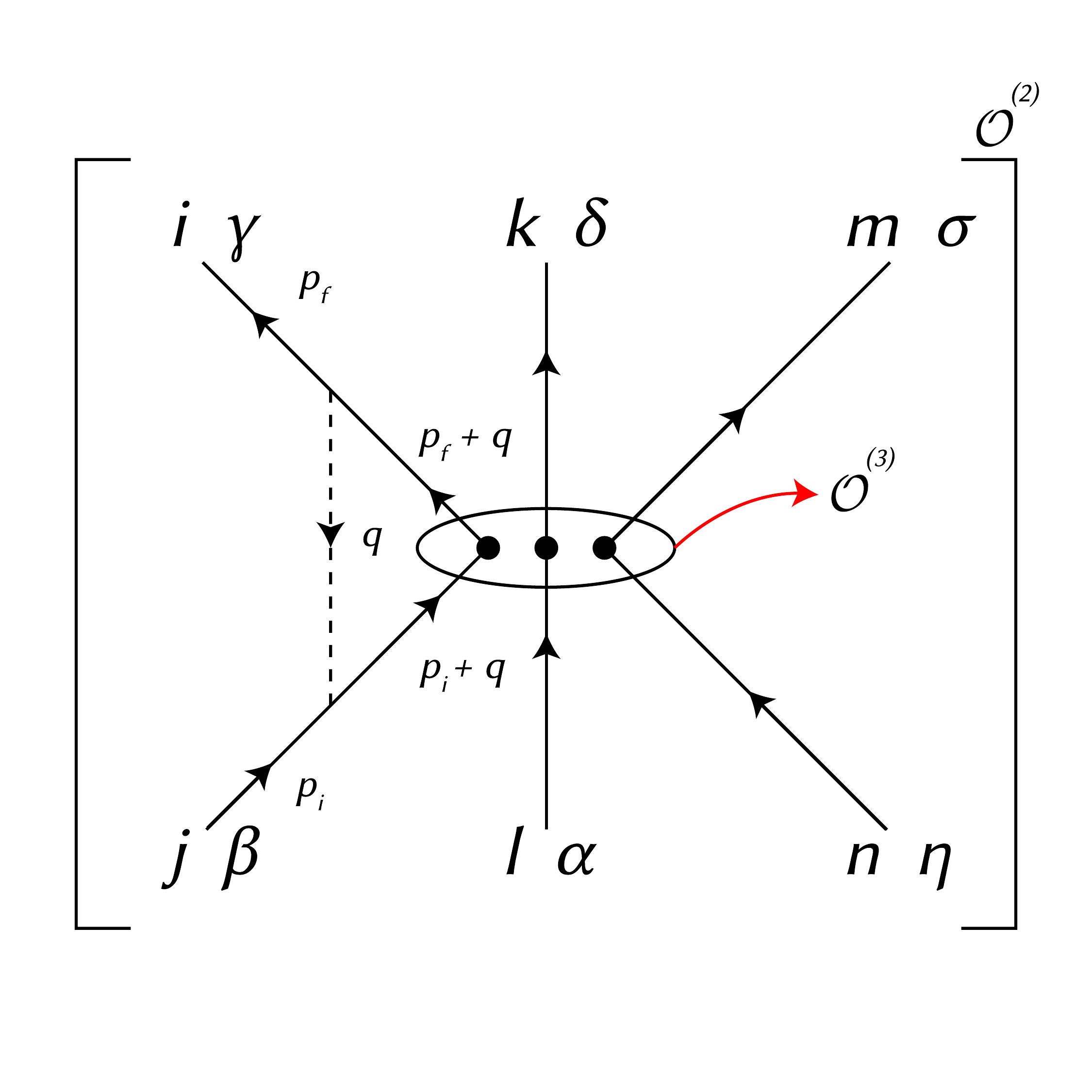}
    \caption{One-loop Yukawa vertex correction.}
    \label{fig:Ap_A_Yukawa_vertex_One_loop}
\end{figure}

The loop integral is given by:
\begin{align}
-iV(p_f, p_i) &= -\widetilde{C}^{(3)}g^2 \int \frac{d^dq}{(2\pi)^d} \frac{i}{q^2 - m^2} \frac{i}{\not{p}_f + \not{q} - M} \frac{i}{\not{p}_i + \not{q} - M}\\
-iV(0, 0)&= -i\widetilde{C}^{(3)}g^2 \int \frac{d^dq}{(2\pi)^d} \frac{1}{q^2 - m^2} \frac{1}{\not{q} - M} \frac{1}{\not{q} - M} \\
&= \widetilde{C}^{(3)}g^2 \frac{\Gamma(\epsilon)}{(4\pi)^{2-\epsilon}} \frac{1}{m^2 - M^2} 
\left[ (m^2)^{1-\epsilon} - (M^2)^{1-\epsilon} \right].
\end{align}

Finally, the vertex counterterm becomes:
\begin{align}
\delta Z^{\phi}_{\mathcal{O}^{(2)}(d3)} &= g^3 \frac{\Gamma(\epsilon)}{(4\pi)^{2-\epsilon}} 
\frac{1}{m^2 - M^2} 
\left[ (m^2)^{1-\epsilon} - (M^2)^{1-\epsilon} \right] \\
&= g^3 \frac{1}{(4\pi)^2} \frac{1}{\epsilon} + \text{finite}.
\end{align}

\section{One-Loop Corrections for $W_L$-Boson, Gluons, and $B^0$-Boson}
\label{sec:AppendixB}

In this appendix, we compute the one-loop corrections for the $W_L$-boson, gluons, and the $B^0$-boson within the general $R_\xi$ gauge. While the procedure remains structurally similar across all cases, the gluons and $B^0$-boson are treated as massless.

\subsection{Renormalization of Gauge Bosons at One Loop}

 The wavefunction renormalization constant is extracted from the self-energy via the standard condition evaluated at zero momentum:
\begin{align}
\frac{\partial \Sigma}{\partial \not{p}} \Big|_{\not{p}=0} &= 0 \quad \Rightarrow \quad \delta Z_\varphi = \frac{\partial \Sigma}{\partial \not{p}} \Big|_{\not{p}=0}. \label{eq:Ap_two_div}
\end{align}

\begin{figure}[h]
    \centering
    \includegraphics[scale=0.2]{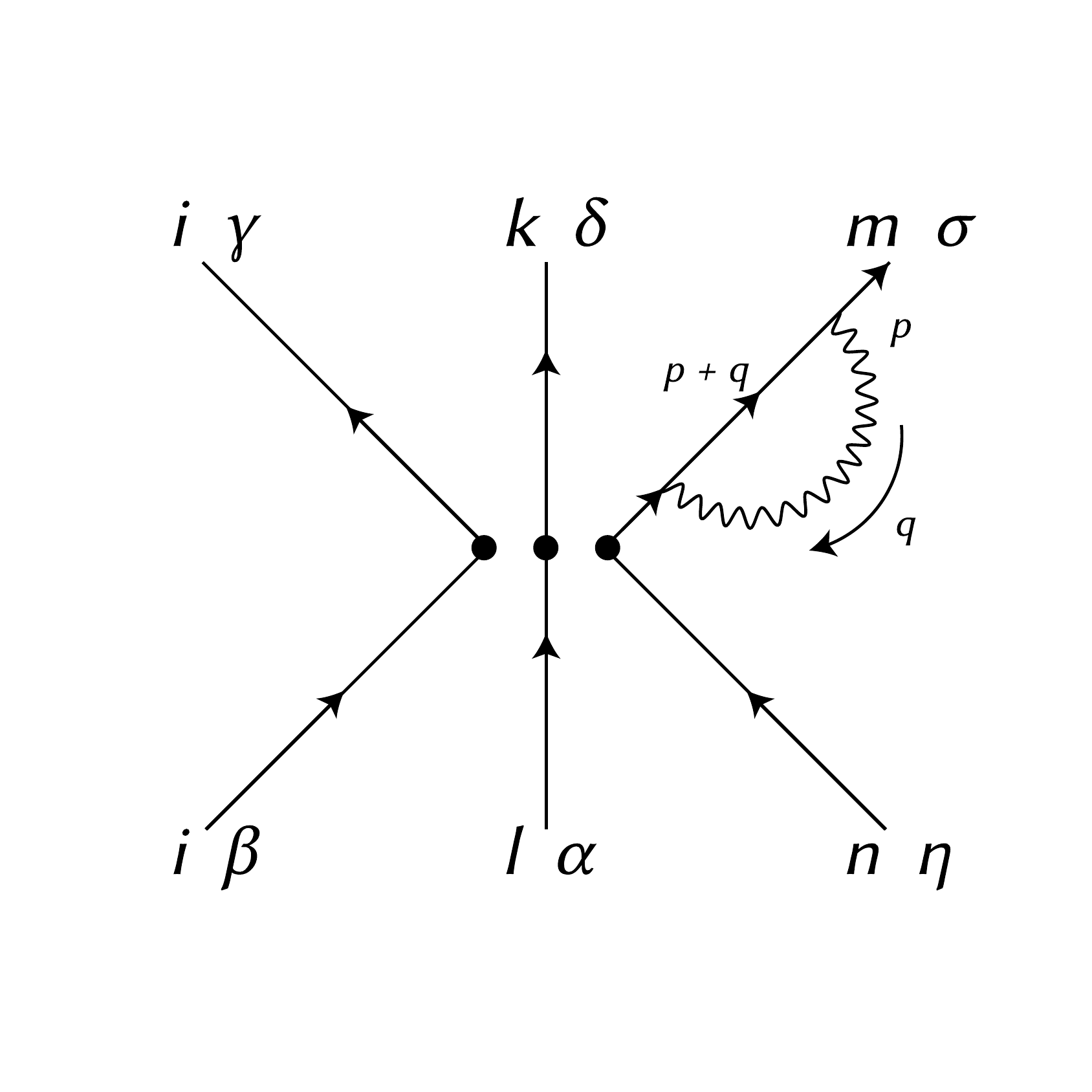}
    \caption{Fermionic one-loop correction to the $W_L$-boson self-energy.}
    \label{fig:Ap_B_W_L_self_One_loop}
\end{figure}

The corresponding Feynman diagram for the $W_L$-boson self-energy, involving a fermion loop, is shown in Fig.~\ref{fig:Ap_B_W_L_self_One_loop}. The boson mass is denoted by $M$, and the fermion mass by $m$.

\vspace{0.7cm}

The amplitude for the self-energy diagram is given by:
\begin{align}
    -i\Sigma(p) = (+i)^2 \int \frac{d^4q}{(2\pi)^4} \frac{(-i)g_{\mu \nu}}{q^2 - M^2 + i\epsilon} \gamma^\mu \frac{i}{\not{p} + \not{q} - m + i\epsilon} \gamma^\nu - \frac{((1-\xi)i)q_\mu q_\nu}{[q^2 - M^2 + i\epsilon]^2} \gamma^\mu \frac{i}{\not{p} + \not{q} - m + i\epsilon} \gamma^\nu \label{eqAp_Two_1_loop_1}
\end{align}

Separating the gauge-independent and gauge-dependent parts, we write:
\begin{align}
    -i\Sigma(p) = -i\Sigma(p)_A +(1-\xi)i\Sigma(p)_B
\end{align}

\textbf{Dirac Algebra Simplification.}\\
The numerator algebra simplifies as:
\begin{align}
    \gamma_\mu(\not{p} + \not{q})\gamma^\nu &= -(\not{p} + \not{q})\gamma_\mu \gamma^\nu + 2(\not{p} + \not{q}) = -(d-2)(\not{p} + \not{q})\\
    m\gamma_\mu \gamma^\nu &= md
\end{align}

\textbf{Dimensional Regularization for Gauge-independent Contribution $\Sigma_A(p)$.}\\
We compute $\Sigma_A(p)$ using dimensional regularization ($d = 4 - \epsilon$):
\begin{align}
    -i\Sigma(p)_A &= -\mu^\epsilon g^2 \int \frac{d^dq}{(2\pi)^d} \frac{1}{q^2 - M^2 + i\epsilon} \gamma_\nu \frac{\not{p} + \not{q} + m}{({p} + {q})^2 - m^2 + i\epsilon} \gamma^\mu \\
    &= -\mu^\epsilon g^2 \int_0^1 dx \left[-(d - 2)\not{p}(1 - x) + m \ d \right] I_{1},
\end{align}

where,
\begin{align}
    I_{1} &= \frac{i}{16\pi^2} \left[\Delta_\epsilon - \ln\left(-p^2 x(1 - x) + m^2 x + M^2 (1 - x)\right)\right]
\end{align}

 and,
\begin{align}
\Delta_\epsilon = \frac{2}{\epsilon} - \gamma + ln 4\pi
\end{align}

Expressing the result in terms of scalar functions:
\begin{align}
\Sigma(p)_A &= A(p^2)_A + B(p^2)_A \not{p}
\end{align}

 where, 
\begin{align}
    A(p^2)_A &= \mu^\epsilon g^2(4- \epsilon)\frac{m}{16\pi^2}\int_0^1 dx \left[\Delta_\epsilon - \ln\left(-p^2 x(1 - x) + m^2 x + M^2 (1 - x)\right)\right]\\
    B(p^2)_A &= -\mu^\epsilon g^2(2- \epsilon)\frac{1}{16\pi^2}\int_0^1 dx (1-x)\left[\Delta_\epsilon - \ln\left(-p^2 x(1 - x) + m^2 x + M^2 (1 - x)\right)\right]
\end{align}

\textbf{Gauge-Dependent Contribution $\Sigma_B(p)$.}\\
Using similar methods, we obtain:
\begin{align}
    -i\Sigma(p)_B &= -\mu^\epsilon g^2 \int \frac{d^dq}{(2\pi)^d} \frac{q^2}{[q^2 - M^2 + i\epsilon]^2} \gamma_\nu \frac{\not{p} + \not{q} + m}{({p} + {q})^2 - m^2 + i\epsilon} \gamma^\mu \\
&= -\mu^\epsilon g^2 \int_0^1 (1 - x) dx \Big[-(d-2)\big[\not{p} (1-x) - 2 \frac{\not{p} x}{d}\big] +m \ d  ] I_{1}\notag \\ 
& + (1-x) [-(d-2)[\not{p}p^2 x^2 (1-x)] + m \ d( p^2 x^2)]I_{2}\Big],
\end{align}

 where,
\begin{align}
I_{1} &= \frac{i}{16\pi^2} \left[\Delta_\epsilon - \ln\left(-p^2 x(1 - x) + m^2 x + M^2 (1 - x)\right)\right]
\end{align}

 and,
\begin{align}
    I_{2} &= \frac{-i}{16\pi^2} \frac{1}{2\left(-p^2 x(1 - x) + m^2 x + M^2 (1 - x)\right)}.
\end{align}

The contribution from the $ \Sigma(p)_B $ can then be written in terms of scalar functions:
\begin{align}
\Sigma(p)_B &= A(p^2)_B + B(p^2)_B \not{p}
\end{align}
\begin{align}
    A(p^2)_B &= \mu^\epsilon g^2(4- \epsilon)\frac{m}{16\pi^2}\int_0^1 (1-x) dx \left[p^2x^2I_2 +I_1\right]\\
    B(p^2)_B &= -\mu^\epsilon g^2\frac{1}{16\pi^2}\int_0^1 (1-x) dx \bigg[-(2- \epsilon)\big[ (1-x) - 2 \frac{ x}{4 -\epsilon}\big]\bigg]I_1 + [- (2-\epsilon)[p^2 x^2 (1-x)]I_2]
\end{align}

\textbf{Full One-Loop Self-Energy and Counterterm.}\\
The full one-loop self-energy is:
\begin{align}
    -i\Sigma(p) &= -i\Sigma(p)_A +(1-\xi)i\Sigma(p)_B \notag\\
    -i\Sigma(p) &= -i[A(p^2)_A + B(p^2)_A \not{p}] +i(1-\xi)[A(p^2)_B + B(p^2)_B \not{p}]\\
    &= -i[A(p^2)_A -(1-\xi)A(p^2)_B] -i[B(p^2)_A -(1-\xi) B(p^2)_B]\not{p}\\
    &= -iA(p^2) -iB(p^2)\not{p}\\
\end{align}

The wavefunction renormalization counterterm is then given as:
\begin{align}
\delta Z_\varphi &= \frac{\partial \Sigma}{\partial \not{p}} \Big|_{\not{p}=0} \notag\\
&= \frac{\partial A}{\partial \not{p}} \Big|_{\not{p}=0} + B + m \frac{\partial B}{\partial \not{p}} \Big|_{\not{p}=0}\\
&= -g^2 \frac{\xi}{(4\pi)^2} \frac{1}{\epsilon} + \text{finite}
\end{align}

\subsection{One-Loop Correction to the $W_L$-Boson Six-Point Vertex}

\begin{figure}[h]
    \centering
    \includegraphics[scale=0.2]{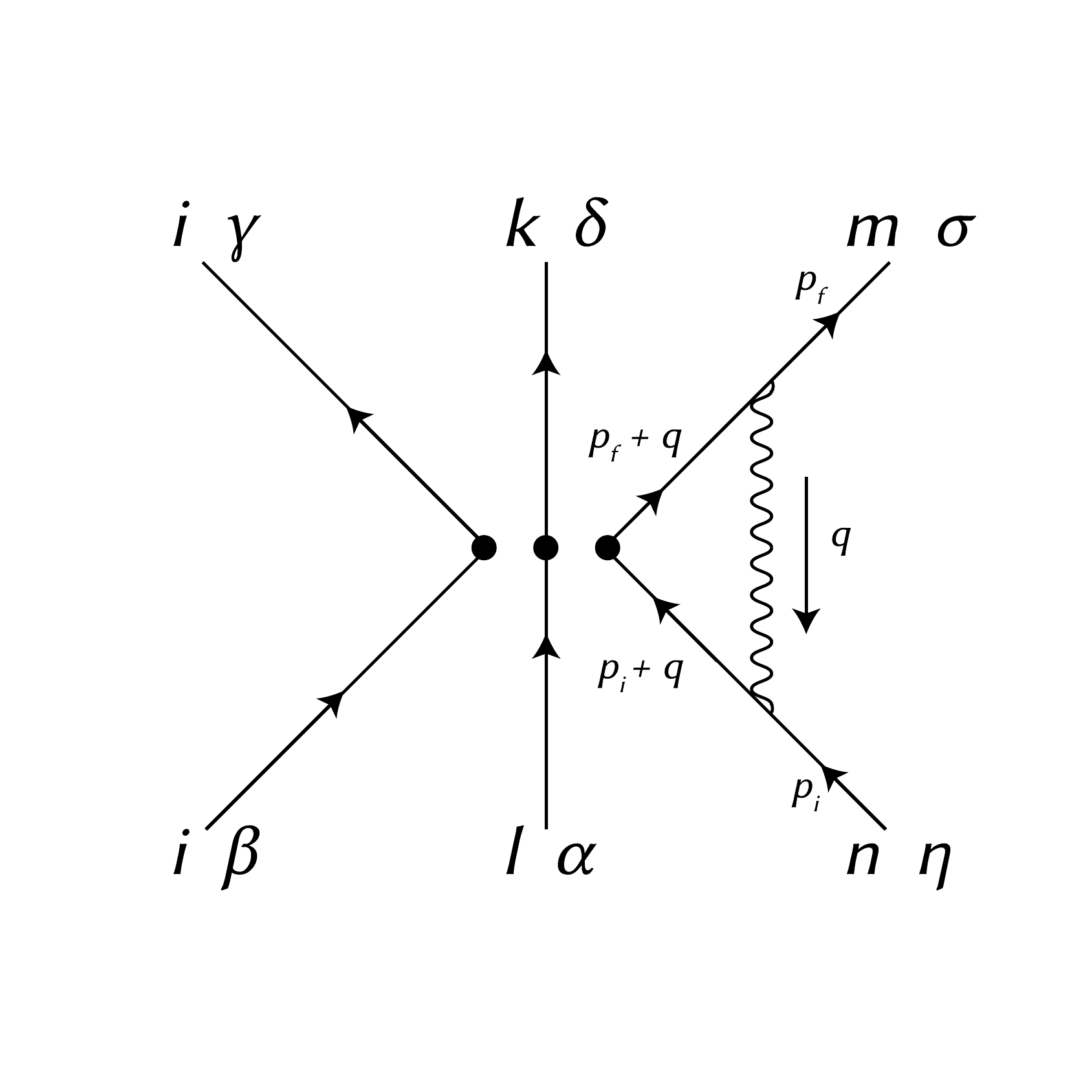}
    \caption{One-loop correction to the $W_L$-boson six-point interaction.}
    \label{fig:Ap_B_W_L_Vertex_One_loop}
\end{figure}

The one-loop correction to the six-point interaction involving the longitudinal component of the $W$-boson is computed in the zero-momentum limit. The relevant Feynman diagram is shown in Fig.~\ref{fig:Ap_B_W_L_Vertex_One_loop}.

\vspace{0.7cm}

The loop-level vertex correction is given by calculating the vertex loop calculation:
\begin{align}
\widetilde{C}^{(2)}g^2(i\mu^{\epsilon/2})^3\Lambda^{\text{loop}}(p_f, p_i) &= \widetilde{C}^{(2)}g^2(i \mu^{\epsilon/2})^3 \notag \\ 
& \times \int \frac{d^dq}{(2\pi)^d} \Bigg[\frac{g_{\rho\sigma}}{q^2 - M^2 + i\epsilon} \nonumber \gamma^\rho \frac{i (\not{q} + \not{p_f} + m')}{(p_f + q)^2 - m'^2 + i\epsilon} \frac{i (\not{q} + \not{p_i}  + m)}{(p_i + q)^2 - m^2 + i\epsilon} \gamma^\sigma\\
&- \frac{((1-\xi)i)q_\rho q_\sigma}{[q^2 - M^2 + i\epsilon]^2} \gamma^\rho \frac{i (\not{q} + \not{p_f} + m')}{(p_f + q)^2 - m'^2 + i\epsilon} \frac{i (\not{q} + \not{p_i}  + m)}{(p_i + q)^2 - m^2 + i\epsilon} \gamma^\sigma\Bigg]
\end{align}

We separate the gauge-independent and gauge-dependent parts of the correction as:

\begin{align}
    \Lambda^{\text{loop}}(p_f, p_i) = \Lambda_A^{\text{loop}}(p_f, p_i) -(1-\xi)\Lambda_B^{\text{loop}}(p_f, p_i)\label{eq:Ap_six_guage}
\end{align}

where $\Lambda^{\text{loop}}$ is related to the full vertex $\Gamma$ through the relation

\begin{align}
i \Gamma(p_f,p_i) &= i({\mathcal{O}}^{(2)} + \Lambda^{\text{loop}}(p_f, p_i) + \delta Z_{\mathcal{O}^(2)})
\end{align}

\textbf{Dimensional Regularization for Gauge-independent Contribution $\Lambda_A^{\text{loop}}(p_f, p_i)$.}\\
We compute $\Lambda_A^{\text{loop}}(p_f, p_i)$ using dimensional regularization ($d = 4 - \epsilon$):

\begin{align}
i \bar{u}(p_f) \Lambda_A^{\text{loop}}(p_f, p_i) u(p_i) &= g^2 \mu^{\epsilon} \int \frac{d^dq}{(2\pi)^d} \frac{\bar{u}(p_f) \gamma_\sigma [ \not{q} + \not{p_f} + m'] [\not{q} + \not{p_i} + m] \gamma^\sigma u(p_i)}{D_0 D_1 D_2}\label{eq:Ap_six_1_loop_1}
\end{align}

where,

\begin{align}
D_0 &= q^2 - M^2 + i\epsilon,  \\
D_1 &= (q + p_f)^2 - m'^2 + i\epsilon,  \\
D_2 &= (q + p_i)^2 - m^2 + i\epsilon. 
\end{align}

\textbf{Dirac Algebra Simplification.}\\
The numerator algebra simplifies as:
\begin{align}
    \gamma_\mu(\not{p}  \not{q})\gamma^\nu &=  4(p\cdot q) - (\not{p}  \not{q})(4-d)\\
    \gamma_\mu(\not{p} + \not{q})\gamma^\nu &= -(\not{p} + \not{q})\gamma_\mu \gamma^\nu + 2(\not{p} + \not{q}) = -(d-2)(\not{p} + \not{q})\\
    m\gamma_\mu \gamma^\nu &= md
\end{align}

\begin{align}
i \bar{u}(p_f) \Lambda_A^{\text{loop}}(p_f, p_i) u(p_i) &= g^2 \mu^{\epsilon} \int \frac{d^dq}{(2\pi)^d} \frac{1}{D_0 D_1 D_2}\bar{u}(p_f) \big[ 4(p^2 +q\cdot(p_f +p)) + dq^2\notag\\
& -(4-d)(\not{p_f}\not{p}+ \not{q}(\not{p_f}+\not{p})) -(d-2)(\not{q}(m +m') + \not{p_f}m + \not{p}m')\notag\\
& +dmm'\big] u(p_i)\label{eq:Ap_six_1_loop_2},
\end{align}

After simplifying the Dirac algebra in the numerator and applying Feynman parameterization and Wick rotation, the integral simplifies to:
\begin{align}
i \bar{u}(p_f) \Lambda_A^{\text{loop}}(p_f, p) u(p_i) &= -ig^2 \mu^{\epsilon} \int_0^1 dx_1 \int_0^{1-x_1} dx_2 \int \frac{d^dq}{(2\pi)^d} \frac{1}{[q^2 + C +i\epsilon]^3}\notag\\
&\times \bar{u}(p_f) \big[ d(A(p_i,p_f)) +B(p_i,p_f) +dq^2 \big] u(p),
\end{align}
 where,

\begin{align}
A(p_i,p_f) &= x^2_1 p_i^2 + x^2_2p_f^2 -2x_1x_2(p_i\cdot{p_f}) + \not{p_f} \not{p_i} + (x_1\not{p_i} -x_2\not{p_f})(\not{p_i}+\not{p_f}) \notag\\
& -[\not{p_i}(m' + x_1(m + m')) + \not{p_f}(m + x_2(m + m'))] + m'm,\\
B(p_i,p_f) &= 4[p_i^2 + (x_1p_i -x_2p_f )\cdot(p_f + p_i)\notag\\
& - (\not{p_f}\not{p_i}) + (x_1\not{p_i} - x_2\not{p_f})\cdot(\not{p_f}+\not{p_i})]  \\
C(p,p_f) &= -x_1M^2 - x_2m'^2 - (1- x_1 -x_2)m^2 + x_2q^2 + x_2p_f^2 + p_i^2 + 2qp_i -x_1p_i^2 - x_2p_f^2\notag\\
& - 2qp_ix_2 - x^2_1p_i^2 - x^2_2p_f^2
\end{align}

After simplifying the integral part, the divergent part yields:
\begin{align}
\Lambda_A^{\text{loop}}(p_f, p) &= -g^2 \mu^{\epsilon} \int_0^1 dx_1 \int_0^{1-x_1} dx_2 \int \frac{d^dq}{(2\pi)^d} \frac{1}{[q^2 + C +i\epsilon]^3} \big[ d(A(p_i,p_f)) +B(p_i,p_f) +dq^2 \big]\\
&= -g^2 \frac{1}{(4\pi)^2} \frac{4}{\epsilon} + \text{finite}
\end{align}

\textbf{Gauge-Dependent Contribution $\Lambda_A^{\text{loop}}(p_f, p)$.}\\
Using similar methods, we obtain:

\begin{align}
i \bar{u}(p_f) \Lambda_B^{\text{loop}}(p_f, p_i) u(p_i) &= g^2 \mu^{\epsilon} \int \frac{d^dq}{(2\pi)^d} \frac{\bar{u}(p_f) q^2 \gamma_\sigma [ \not{q} + \not{p_f} + m'] [\not{q} + \not{p_i} + m] \gamma^\sigma u(p_i)}{D_0^2 D_1 D_2},
\end{align}

 where,
\begin{align}
D_0 &= q^2 - M^2 + i\epsilon, \label{eq:Ap_six_2_D_0} \\
D_1 &= (q + p_f)^2 - m'^2 + i\epsilon, \label{eq:Ap_six_2_D_1} \\
D_2 &= (q + p_i)^2 - m^2 + i\epsilon. \label{eq:Ap_six_2_D_2}
\end{align}

\textbf{Dirac Algebra Simplification.}\\
The numerator algebra simplifies as:
\begin{align}
i \bar{u}(p_f) \Lambda_B^{\text{loop}}(p_f, p_i) u(p_i) &= g^2 \mu^{\epsilon} \int \frac{d^dq}{(2\pi)^d} \frac{q^2}{D_0^2 D_1 D_2}\bar{u}(p_f) \big[ 4(p_i^2 +q\cdot(p_f +p_i)) + dq^2\notag\\
& -(4-d)(\not{p_f}\not{p_i}+ \not{q}(\not{p_f}+\not{p_i})) -(d-2)(\not{q}(m +m') + \not{p_f}m + \not{p_i}m')\notag\\
& +dmm'\big] u(p_i),
\end{align}

Applying Feynman parameterization and Wick rotation, we obtain:
\begin{align}
i \bar{u}(p_f) \Lambda_B^{\text{loop}}(p_f, p_i) u(p_i) &= -ig^2 \mu^{\epsilon} \int_0^1 dx_1 \int_0^{1-x_1} dx_2 \int \frac{d^dq}{(2\pi)^d} \frac{(1-x)[q^2 + x^2_1 p_i^2 + x^2_2p_f^2 -2x_1x_2p_i\cdot{p_f}]}{[q^2 + C +i\epsilon]^4}\notag\\
& \times\bar{u}(p_f) \big[ d(A(p_i,p_f)) +B(p_i,p_f) +dq^2 \big] u(p_i)\\
&= -ig^2 \mu^{\epsilon} \int_0^1 dx_1 \int_0^{1-x_1} dx_2 \int \frac{d^dq}{(2\pi)^d} \Bigg[ \frac{\bar{u}(p_f) \big[ dq^4+ q^2(dA + dH +B) \big] u(p_i)}{[q^2 + C +i\epsilon]^4}\notag\\
& + \frac{\bar{u}(p_f)\big[ H(dA + B) \big]u(p_i)}{[q^2 + C +i\epsilon]^4}\Bigg]
\end{align}

 where,
\begin{align}
A(p_i,p_f) &= x^2_1 p_i^2 + x^2_2p_f^2 -2x_1x_2p_i\cdot{p_f} + \not{p_f} \not{p_i} + (x_1\not{p_i} -x_2\not{p_f})(\not{p_i}+\not{p_f}) \notag\\
& -[\not{p_i}(m' + x_1(m + m')) + \not{p_f}(m + x_2(m + m'))] + m'm,\\
B(p_i,p_f) &= 4[p_i^2 + (x_1p_i -x_2p_f )\cdot(p_f + p_i)\notag\\
& - (\not{p_f}\not{p_i}) + (x_1\not{p_i} - x_2\not{p_f})\cdot(\not{p_f}+\not{p_i})]  \\
C(p_i,p_f) &= -x_1M^2 - x_2m'^2 - (1- x_1 -x_2)m^2 + x_2q^2 + x_2p_f^2 + p_i^2 + 2qp_i -x_1p_i^2 - x_2p_f^2\notag\\
& - 2qp_ix_2 - x^2_1p_i^2 - x^2_2p_f^2 \\
H(p_i,p_f) &= x^2_1 p_i^2 + x^2_2p_f^2 -2x_1x_2p_i\cdot{p_f}
\end{align}

After simplifying the integral part, the divergent part yields:
\begin{align}
\Lambda_B^{\text{loop}}(p_f, p_i)  &= -g^2 \mu^{\epsilon}\int_0^1 dx_1 \int_0^{1-x_1} dx_2 \int \frac{d^dq}{(2\pi)^d} \frac{ (1-x)\big[ dq^4+ q^2(dA + dH +B) + H(dA + B) \big]}{[q^2 + C +i\epsilon]^4}\\
&= -g^2 \frac{1}{(4\pi)^2} \frac{1}{\epsilon} + \text{finite}
\end{align}

\textbf{Full One-Loop Six-Point Vertex and Counterterm.}\\
Substituting both components into Eq.~\ref{eq:Ap_six_guage}, the total divergence becomes:
\begin{align}
    \Lambda^{\text{loop}}(p_f, p_i) &= (-g^2 \frac{1}{(4\pi)^2} \frac{4}{\epsilon} + \text{finite}) -(1-\xi)(-g^2 \frac{1}{(4\pi)^2} \frac{1}{\epsilon} + \text{finite})\\
    &= -g^2 \frac{(3+\xi)}{(4\pi)^2} \frac{1}{\epsilon} + \text{finite}
\end{align}
The counter term is calculated at $\not{p_i},\not{p_l} = 0$
\begin{align}
Z_{\mathcal{O}^{(2)}} &= -\Lambda^{\text{loop}}(p_f, p_i) \Big|_{\not{p_i},\not{p_f}=0}\\
& = g^2 \frac{(3+\xi)}{(4\pi)^2} \frac{1}{\epsilon} + \text{finite}
\end{align}

\bibliographystyle{JHEPCust}
\bibliography{vecB}

\end{document}